\documentclass[12pt]{article}%
\pdfoutput=1
\usepackage[nosort]{cite}
\usepackage{graphicx}
\usepackage{multicol}
\usepackage{amsfonts}
\usepackage{amssymb}
\usepackage{amsmath}
\usepackage{heck}
\usepackage{afterpage}
\usepackage{setspace}
\usepackage{verbatim}
\usepackage{color}
\usepackage{slashed}
\usepackage{longtable}
\usepackage{float}
\usepackage{subcaption}
\usepackage{epsfig}
\usepackage{enumerate}
\usepackage{epstopdf}
\usepackage[enableskew, vcentermath]{youngtab}
\usepackage{adjustbox}
\newsavebox{\mysavebox}

\usepackage{youngtab}
\usepackage{tikz}
\usepackage[margin=1in]{geometry}
\usepackage{titletoc}%
\usepackage{hyperref}
\hypersetup{colorlinks,%
citecolor=black,%
filecolor=black,%
linkcolor=black,%
urlcolor=black}
\providecommand{\U}[1]{\protect\rule{.1in}{.1in}}
\usetikzlibrary{decorations.markings}

\numberwithin{equation}{section}

\usepackage{mathtools}
\usepackage{pgfplots}
\pgfplotsset{compat=1.3}
\usetikzlibrary{chains, external}
\usepgfplotslibrary{external}
\tikzset{node distance=2em, ch/.style={circle,draw,on chain,inner sep=2pt},chj/.style={ch,join},every path/.style={shorten >=4pt,shorten <=4pt},line width=1pt,baseline=-1ex}

\newcommand{\CO}{\mathcal{O}}
\newcommand{\bbP}{\mathbb{P}}
\newcommand{\bbF}{\mathbb{F}}
\newcommand{\ft}{\tilde{f}}
\newcommand{\gt}{\tilde{g}}

\tikzstyle{startstop} = [rectangle, rounded corners, minimum width=3cm, minimum height=1cm,text centered, draw=black, fill=blue!10]
\tikzstyle{startstop} = [rectangle, rounded corners, minimum width=3cm, minimum height=1cm,text centered, draw=black, fill=blue!10]

\tikzstyle{io} = [trapezium, trapezium left angle=70, trapezium right angle=110, minimum width=3cm, minimum height=1cm, text centered, draw=black, fill=blue!30]

\tikzstyle{process} = [rectangle, minimum width=3cm, minimum height=1cm, text centered, draw=black, fill=orange!30]
\tikzstyle{decision} = [diamond, minimum width=3cm, minimum height=1cm, text centered, draw=black, fill=green!30]

\tikzstyle{arrow} = [thick,->,>=stealth]

\tikzset{->-/.style={decoration={
  markings,
  mark=at position #1 with {\arrow[scale=2.4]{>}}},postaction={decorate}}}

\makeatletter \@addtoreset{equation}{section} \makeatother

\begin{document}

\date{June 2018}

\title{Towards Exotic Matter and\\[4mm] Discrete Non-Abelian Symmetries in F-theory}

\institution{UPENN}{\centerline{${}^{1}$Department of Physics and Astronomy, University of Pennsylvania, Philadelphia, PA 19104, USA}}

\institution{MARIBOR}{\centerline{${}^{2}$Center for Applied Mathematics and Theoretical Physics, University of Maribor, Maribor, Slovenia}}

\authors{Mirjam Cveti\v{c}\worksat{\UPENN, \MARIBOR}\footnote{e-mail: {\tt cvetic@physics.upenn.edu}},
Jonathan J. Heckman\worksat{\UPENN}\footnote{e-mail: {\tt jheckman@sas.upenn.edu}}
and Ling Lin\worksat{\UPENN}\footnote{e-mail: {\tt lling@physics.upenn.edu}}}

\abstract{We present a prescription in F-theory for realizing matter in ``exotic'' representations of
product gauge groups. For 6D vacua, bifundamental hypermultiplets
are engineered by starting at a singular point in moduli space which includes
6D superconformal field theories coupled to gravity. A deformation in Higgs branch moduli space takes us to
a weakly coupled gauge theory description. In the corresponding elliptically fibered Calabi--Yau threefold,
the minimal Weierstrass model parameters $(f,g,\Delta)$ vanish at collisions of the discriminant
at least to order $(4,6,12)$, but with sufficiently high order of tangency to ensure the existence of T-brane deformations
to a weakly coupled gauge theory with exotic bifundamentals. We present explicit examples including
bifundamental hypermultiplets of $\mathfrak{e}_7 \times \mathfrak{su}_2$
and $\mathfrak{e}_6 \times \mathfrak{su}_3$, each of which have dual heterotic orbifold descriptions.
Geometrically, these matter fields are delocalized across multiple points of an F-theory geometry.
Symmetry breaking with such representations can be used to produce high dimension representations of simple gauge groups
such as the four-index symmetric representation of $\mathfrak{su}_2$ and the three-index symmetric representation
of $\mathfrak{su}_3$, and after further higgsing can yield discrete non-abelian symmetries.}

{\small \texttt{\hfill UPR-1290-T}}

\maketitle

\tableofcontents

\enlargethispage{\baselineskip}

\setcounter{tocdepth}{2}

\newpage

\section{Introduction \label{sec:INTRO}}

One of the important lessons from string theory is that not all
quantum field theories can be consistently coupled to quantum gravity.  From a bottom up
perspective, string theory can be viewed as pointing the way to subtle consistency
conditions which might otherwise be missed. With this in mind,
it is clearly important to delineate the full space of quantum field theories
which can be realized in string theory \cite{Vafa:2005ui, Ooguri:2006in}.

For example, unlike from a purely effective field theory perspective, there seem to be clear limitations on the matter content in
weakly coupled (supersymmetric) gauge theories that are realized in string compactifications. 
The most straightforward examples in perturbative open string theories involve matter in various two-index representations of the classical gauge groups $U(N)$, $SO(N)$, $Sp(N)$. This can be widened in F-theory \cite{Vafa:1996xn, Morrison:1996na, Morrison:1996pp} to include all the exceptional
groups, but again, matter fields tend to transform in low dimension representations (namely the fundamental or small extensions thereof)
such as the $\mathbf{27}$ of $E_6$.

It is sometimes possible to arrange for more exotic representations in various free fermion constructions, as for example with heterotic strings at higher Kac--Moody levels (see, e.g., \cite{Dienes:1996du} and references therein), but even here, there are limitations to what can be obtained. The question of which 6D supergravity models can be embedded in F-theory has been the subject of extensive research \cite{Kumar:2009ac, Kumar:2010ru, Park:2011wv, Anderson:2015cqy, Cvetic:2017epq, Klevers:2017aku, Taylor:2018khc}.

A related but far less explored question has to do with the structure of possible discrete gauge groups which can be consistently
realized. There are by now various examples of models with abelian symmetries, but even in this case the order of the group
is typically quite small. Even fewer examples are available for
non-abelian discrete groups. The latter are of interest in their own right, and also have applications to topological
field theory and particle physics. Some work on the stringy realization of discrete non-abelian groups
includes \cite{Gukov:1998kn, Kobayashi:2006wq, Nilles:2012cy, BerasaluceGonzalez:2012vb, BerasaluceGonzalez:2012zn, Braun:2017oak} as well as \cite{Grimm:2015ona} in F-theory.
As of this writing there does not appear to be a systematic method to determine whether a particular
finite group can appear.

In a certain sense, the appearance of matter in high dimension representations of continuous gauge groups and non-abelian discrete gauge
groups are related since the latter can be obtained by giving vevs to scalars transforming in high dimension representations \cite{Frampton:1994rk}. From this perspective, it is natural to ask whether there are systematic ways to find compactifications which contain such
structures.

Part of the issue with realizing higher dimension representations is that the standard way to engineer
low energy effective field theories via string compactification typically breaks a substantial part
of the higher-dimensional gauge group, leaving only a lower rank remnant behind.
For example, in the context of compactifications of the heterotic string,
one often specifies a vector bundle with structure group $K \subset E_8$. The commutant subgroup $G$
is then all that survives in the low energy theory. This in turn limits the matter content $G$ to low dimension
representations. Related behavior occurs in F-theory and M-theory models, where the appearance of a non-trivial
vector bundle is instead signalled by the presence of a generic unfolding of a singularity.

Tuning the moduli of a compactification provides one way to realize such exotica.
For example, it has been known since the early days of heterotic strings compactified on orbifolds \cite{Dixon:1985jw, Dixon:1986jc},
that it is possible to achieve matter fields in the $(\mathbf{56},\mathbf{2})$ of $\mathfrak{e}_7 \times \mathfrak{su}_2$ as well as the $(\mathbf{27},\mathbf{3})$ of $\mathfrak{e}_6 \times \mathfrak{su}_3$ (see e.g. \cite{Honecker:2006qz} for this and related examples).
Perhaps surprisingly, in F-theory the corresponding geometric realization of these exotic bifundamental matter fields has only been studied in a
few cases (see, e.g., \cite{Ludeling:2014oba}). Other examples with exotic matter representations in F-theory involve
matter on a singular divisor \cite{Klevers:2017aku, Cvetic:2015ioa, Klevers:2016jsz,  Raghuram:2017qut}. As far as we are aware, however, there is no systematic
``rule of thumb'' for how to realize such matter representations in F-theory.

With this in mind, our aim in this work will be to develop a set of geometric criteria for engineering
exotic matter field representations in F-theory. In field theory terms, our main interest will be on 6D theories in flat space with hypermultiplets in ``exotic'' representations $(\mathbf{R}_1 , \mathbf{R}_2)$ of some gauge algebra $\mathfrak{g}_1 \times \mathfrak{g}_2$.
As already mentioned, activating vevs for these matter fields also provides a mechanism for
generating examples of discrete non-abelian symmetries.

From a geometric perspective, we consider elliptically fibered Calabi--Yau threefolds $X \rightarrow B$
with minimal Weierstrass model:
\begin{equation}
y^2 = x^3 + fx + g
\end{equation}
so that the vanishing locus for the discriminant $\Delta = 4f^3 + 27g^2$ has components supporting gauge
algebra $\mathfrak{g}_1 \times \mathfrak{g}_2$. Matter
in low-dimensional representations comes about from collisions of these components.
We get weakly coupled matter when the multiplicity of vanishing for $(f,g,\Delta)$ is below $(4,6,12)$.
We can also realize 6D superconformal field theories (SCFTs) when the multiplicity of vanishing is at least $(4,6,12)$ since then,
the elliptic fiber does not remain in Kodaira--Tate form \cite{Morrison:1996na, Morrison:1996pp, Bershadsky:1996nu, Aspinwall:1997ye, Heckman:2013pva} (see \cite{Heckman:2018jxk} for a recent review of 6D SCFTs
from a top down perspective). In this case, to truly understand the physics of the F-theory model, we need to perform blow-ups
of the base. Collapsing the additional $\mathbb{P}^1$'s down to zero size then takes us
back to the SCFT point. An interesting open problem is to determine which 6D SCFTs can be consistently coupled to
gravity. For some results in this direction, see e.g. \cite{DelZotto:2014fia, Anderson:2015cqy, Anderson:2018heq}.

Now, in the context of F-theory, when one attempts to use the standard Tate algorithm to
engineer ``exotic'' high dimension representations $(\mathbf{R}_1 , \mathbf{R}_2)$
of some gauge algebra $\mathfrak{g}_1 \times \mathfrak{g}_2$,
one inevitably finds that the multiplicity of vanishing for $(f,g,\Delta)$ is at least $(4,6,12)$,
indicating the appearance of a 6D SCFT lurking at some point of moduli space.

\begin{figure}[t!]
\begin{center}
\scalebox{1}[1]{
\includegraphics[trim={2cm 3cm 0cm 4cm},clip,scale=0.5]{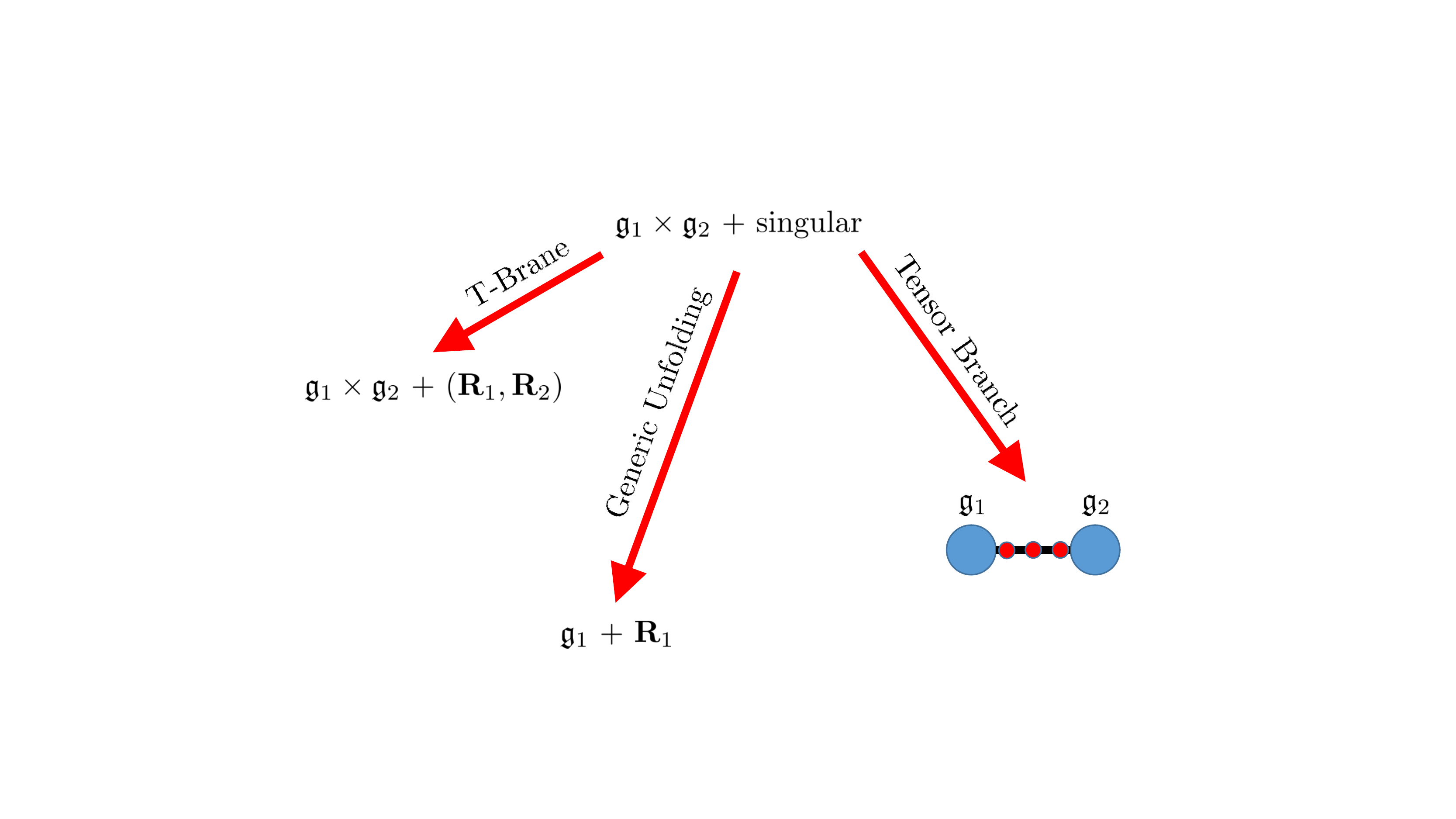}}
\end{center}
\caption{Depiction of the deformations of singular F-theory models which can generate an exotic bifundamental. The minimal
Weierstrass model contains points where the multiplicity of vanishing for $(f,g,\Delta)$ is at least $(4,6,12)$. T-brane
deformations lead to models with exotic bifundamentals, and tensor branch deformations lead to deformations of 6D SCFTs
coupled to gravity.}
\label{fig:GENERALFLOWS}
\end{figure}

But as has been appreciated for some time, the Weierstrass model provides only partial
information on the effective field theory specified by an F-theory model. To remove possible ambiguities
in reading off the low energy effective field theory, it is necessary to indicate a \textit{neighborhood} in moduli space,
and in particular more than just a singular Weierstrass model. At a smooth point of moduli space, this involves indicating the value of the complex structure moduli as well as the intermediate Jacobian of the threefold \cite{Aspinwall:1998he, Anderson:2013rka, Anderson:2017rpr}, and
in singular limits, the moduli space is completed by T-brane configurations \cite{Anderson:2013rka, Anderson:2017rpr} (see also \cite{Cecotti:2010bp, Collinucci:2014taa, Collinucci:2014qfa, Bena:2016oqr, Marchesano:2016cqg, Bena:2017jhm, Marchesano:2017kke}). This is particularly important in the case of $(4,6,12)$ points because motion in a T-brane direction means one is not at a singular point of the effective field theory moduli space. See figure \ref{fig:GENERALFLOWS} for a depiction of the different deformations of a
singular F-theory model.

When the collision of divisors supporting algebras $\mathfrak{g}_1$ and $\mathfrak{g}_2$ occurs with
a sufficiently high order of tangency, additional moduli are also often localized. This in
turn leads to the appearance of additional deformations with a weakly coupled interpretation.
From a bottom up perspective, a necessary condition for this to occur is that we can match quantities such as the gravitational anomalies of the theory with 6D conformal matter \cite{DelZotto:2014hpa, Heckman:2014qba, Heckman:2015bfa}
with those at the weakly coupled point. From a top down perspective, we can
provide checks on our proposal by applying heterotic/F-theory duality in the special limit where the heterotic string propagates on an
orbifold.

In this work, we focus on two canonical examples where this phenomenon occurs, though it is clear our method generalizes.
First, we consider exotic bifundamental matter in the $(\mathbf{56},\mathbf{2})$ of $\mathfrak{e}_7 \times \mathfrak{su}_2$.
This arises in compactifications of heterotic strings on the orbifold $T^4 / \mathbb{Z}_2$, and the F-theory dual description of this case has been studied for example in \cite{Ludeling:2014oba}. Here, we clarify some aspects of this construction, in particular the structure of the moduli space on the tensor branch of the associated SCFTs. We then turn to exotic bifundamental matter in the $(\mathbf{27},\mathbf{3})$ of $\mathfrak{e}_6 \times \mathfrak{su}_3$ which arises in compactifications of heterotic strings on the orbifold $T^4 / \mathbb{Z}_3$. In this case, there are additional subtle features in the global F-theory model, since the full gauge algebra is $\mathfrak{e}_6 \times \mathfrak{su}_3 \times \mathfrak{u}(1) \times \mathfrak{e}_7$.
Both examples have the peculiar feature that a single bifundamental hypermultiplet is spread across several points of the F-theory base.
As such, their appearance cannot be a purely local effect, as it is for standard F-theory matter, and has to be a consequence of the global nature of our models.

The resulting 6D effective field theories can be further taken as the starting point for realizing additional ``exotic structures'' in F-theory by further higgsing. Strictly speaking, the compact models we present contain only a single exotic bifundamental, whereas to initiate further breaking patterns it is necessary to have at least two such hypermultiplets. This appears
possible to arrange in various local geometries, as well as 4D models, and suggests a rich class of additional exotic structures.

For example, breaking patterns involving the $(\mathbf{56},\mathbf{2})$ of $\mathfrak{e}_7 \times \mathfrak{su}_2$ can realize an $\mathfrak{su}_2$ gauge theory with matter in the $\mathbf{5}$ of $\mathfrak{su}_2$, which is higher than the $\mathbf{4}$ of $\mathfrak{su}_2$ realized via matter on a singular divisor (see reference \cite{Klevers:2017aku}).
In the case of the $(\mathbf{27},\mathbf{3})$ of $\mathfrak{e}_6 \times \mathfrak{su}_3$, a further higgsing yields an $\mathfrak{su}_3$ gauge theory with matter in the three-index symmetric representation ($\mathbf{10}$) of $\mathfrak{su}_3$, thus exceeding the dimension of the two-index symmetric ($\mathbf{6}$) constructed in \cite{Cvetic:2015ioa}.
Proceeding even further, activating vevs for these high dimension representations leads us to finite non-abelian gauge groups.
In the case of $\mathfrak{su}_2$ breaking via a vev for the $\mathbf{5}$ of $\mathfrak{su}_2$ (see, e.g., \cite{Adulpravitchai:2009kd}), this yields the quaternion group, and in the case of $\mathfrak{su}_3$ breaking via vevs for the $\mathbf{10}$ as well $\mathbf{6}$ of $\mathfrak{su}_3$ (see, e.g., \cite{Luhn:2011ip}), this yields $A_4$, the alternating group on four letters.

The rest of this paper is organized as follows. In section \ref{sec:PRESCRIPTION} we present our proposal for connecting points with ``conformal matter singularities'' in the presence of higher tangencies of the Weierstrass model with the existence of weakly coupled points in the moduli space. In section \ref{sec:E7SU2} we turn to the F-theory realization of the $(\mathbf{56},\mathbf{2})$ of $\mathfrak{e}_7 \times \mathfrak{su}_2$, and in section \ref{sec:E6SU3} we analyze the F-theory realization of the $(\mathbf{27},\mathbf{3})$ of $\mathfrak{e}_6 \times \mathfrak{su}_3$. Section \ref{sec:BREAKING} discusses breaking patterns using exotic bifundamentals. We conclude in section \ref{sec:CONC} and also discuss some future areas of potential investigation. Some additional technical elements are deferred to the Appendices.

\section{Prescription for Exotic Bifundamentals \label{sec:PRESCRIPTION}}

In this section we present a general prescription for constructing examples in F-theory
of matter in ``exotic'' bifundamental representations. The main setup we wish to consider involves collisions of
two components of the discriminant locus, supporting respective gauge algebras $\mathfrak{g}_1$ and $\mathfrak{g}_2$.
At the collision point, matter will be localized. In terms of the minimal Weierstrass model:
\begin{equation}
y^2 = x^3 + fx + g,
\end{equation}
the discriminant locus $\Delta = 4f^3 + 27g^2$ factorizes into distinct components:
\begin{equation}
\Delta = \Delta_1 \Delta_2 \cdots.
\end{equation}
Geometrically, the elliptically fibered Calabi--Yau is ill-defined when the multiplicity of vanishing for $(f,g,\Delta)$ at a point of the base
is at least $(4,6,12)$. In this case, blow-ups of the base are necessary, and the physical interpretation is the appearance of a 6D superconformal field theory with degrees of freedom localized at this singular point of the geometry \cite{Morrison:1996na, Morrison:1996pp, Bershadsky:1996nu, Aspinwall:1997ye, Heckman:2013pva}. When there is a collision but the multiplicity of vanishing is less singular, we instead expect weakly coupled hypermultiplets. For a recent review on the construction of 6D SCFTs in F-theory, see \cite{Heckman:2018jxk}.

Following the standard Tate algorithm \cite{Bershadsky:1996nu, Katz:2011qp}, we find that engineering bifundamentals for gauge algebras such as $\mathfrak{e}_7 \times \mathfrak{su}_2$ or $\mathfrak{e}_6 \times \mathfrak{su}_3$ inevitably leads to a singular collision point where the elliptic fibers do not remain in Kodaira--Tate form. For example, the collision of a type III$^\ast$ and a type III fiber can, in local coordinates $(t,u)$ of the base, be written as:
\begin{equation}
y^2 = x^3 + t u^3 x + t^2 u^5,
\end{equation}
so to interpret this in terms of a Calabi--Yau geometry, we must blow-up the base at the point $t = u = 0$.
By a similar token, the collision of a type IV$^\ast$ and a split type IV fiber can be written as:
\begin{equation}
y^2 = x^3 + t^2 u^3 x + t^2 u^4,
\end{equation}
which again suggests the need for a blow-up at the collision point $t = u = 0$. The analysis of the resulting
F-theory geometries was first carried out in \cite{Bershadsky:1996nu, Aspinwall:1997ye}, and was subsequently systematized
in later work on 6D SCFTs (see \cite{Heckman:2018jxk} for a recent review).

But the physical theory involves additional data beyond that specified by the minimal Weierstrass model. Indeed,
in the context of 6D vacua, the Higgs branch of F-theory vacua at a smooth point in moduli space
is specified by the complex structure moduli as well as the (Weil) intermediate Jacobian of the Calabi--Yau threefold.
In singular limits of complex structure moduli \cite{Aspinwall:1997ye, Anderson:2013rka, Anderson:2017rpr},
the moduli of the intermediate Jacobian are interpreted as T-brane data \cite{Cecotti:2010bp, Collinucci:2014taa,
Collinucci:2014qfa, Bena:2016oqr, Marchesano:2016cqg, Bena:2017jhm, Marchesano:2017kke}.
From this perspective, a seemingly singular presentation of the Calabi--Yau geometry may actually reside at a non-singular point of
moduli space in the physical theory.

An illustrative example of this kind is the F-theory dual for heterotic strings compactified on a smooth K3 surface with vector bundle specified by the standard embedding, that is, we embed the spin connection in one of the $E_8$ factors. The F-theory geometry is the same as the one we would specify for a theory of $24$ small $E_8$ instantons \cite{Aspinwall:1997ye}. We have in both cases (again using local coordinates):
\begin{equation}
y^2 = x^3 + g_{24}(t) u^5 + f_8(t) u^4 x + g_{12}(t) u^6 + ....
\end{equation}
Geometrically, each zero of $g_{24}$ specifies a type II fiber, and the collision with the $E_8$ locus (a type II$^\ast$ fiber) over $u=0$
indicates a superconformal field theory, provided we are at the origin of the moduli space \cite{Morrison:1996na, Morrison:1996pp}.

From a physical perspective, we can start at the singular point in moduli space associated with codimension four instantons of
the ten-dimensional $E_8$ gauge theory, in which case the zeros of $g_{24}$ indicate the locations of these instantons. Dissolving these
instantons into flux (by making the instantons have finite size) moves us onto the Higgs branch of the moduli space, and
takes us to a non-singular point in moduli space which is smoothly connected to the heterotic tangent bundle description. At this point
in moduli space, we have a standard weakly coupled interpretation of the gauge theory and matter content given by an $\mathfrak{e}_7$ gauge
theory with $20$ half-hypermultiplets in the fundamental representation.

We can also coalesce multiple instantons at the same location. For example, the local presentation:
\begin{equation}
y^2 = x^3 + u^5 (u + \alpha t^k) + ...
\end{equation}
involves a higher order tangency of intersection at $t = u = 0$, and is associated with $k$ small instantons at the same locus.
The appearance of more instantons at the same location of the geometry means there are additional moduli are available. Again,
in heterotic terms we can dissolve these instantons as flux inside the $E_8$ locus, leading to a weakly coupled point of moduli space.
In these cases as well, T-brane deformations are available which leave the Weierstrass model unchanged.

Putting these facts together, we see that even when we encounter a singular point where the
multiplicity of vanishing for $(f,g,\Delta)$ is at least
$(4,6,12)$, it may still be possible to move in Higgs branch moduli space whilst still remaining at a singular
point of complex structure moduli. Moreover, we see that by increasing the order of tangency for the collision of components of the discriminant, we have additional moduli.
These additional moduli mean there can be further T-brane deformations available to move us to a weakly coupled description.
Note that for heterotic strings compactified on a smooth K3 surface, we need at least four small instantons to participate in the breaking
pattern to $E_7$.
The standard way to check whether such a Higgs branch transition is possible is to match various anomalies between the tensor branch and Higgs branch descriptions.
In particular, we need to ensure that there is no change in the contribution to the gravitational anomaly:
\begin{equation}
\delta H - \delta V + 29 \delta T = 0,
\end{equation}
where $\delta H$ denotes the change in the number of hypermultiplets, $\delta V$ the change in the
number of vector multiplets, and $\delta T$ the change in the number of tensor multiplets.

Taking into account these considerations, we see that even when we encounter collisions of the discriminant locus which
do not remain in Kodaira--Tate form, the physical theory may nevertheless be described by a weakly coupled model.
From this perspective, to engineer examples of ``exotic'' bifundamental representations, we apply the following procedure:
\begin{itemize}
\item Engineer a collision between two components of the discriminant with gauge algebra $\mathfrak{g}_1 \times \mathfrak{g}_{2}$.
\item Determine the tensor branch by blowing up the collision point, using the algorithm outlined in references \cite{Bershadsky:1996nu, Aspinwall:1997ye} (see also \cite{Heckman:2013pva}).
\item Passing back to the origin of the tensor branch, increase the order of tangency to generate a higher dimension for the Higgs branch.
\item Introduce a candidate weakly coupled gauge theory with bifundamentals with the same gravitational anomaly as the tensor branch theory.
\item When available, use a dual heterotic description as a cross check.
\end{itemize}
In practice, we find that exotic bifundamentals are delocalized across multiple collision points.
While it might indeed appear exotic in F-theory, in the hetorotic dual 
the exotic bifundamental comes from the untwisted sector. This also illustrates that 
some aspects of exotic bifundamentals are not completely local.

As we have already noted, the step of tuning to increase the order of tangency is sometimes necessary for a T-brane
deformation to exist at all. It would be interesting to determine from first principles the minimal order of tangency necessary
to accommodate exotic bifundamentals, but in this work we shall confine our analysis to a recursive ``trial and error'' procedure.

\section{Engineering the $(\mathbf{56} , \mathbf{2})$ of $\mathfrak{e}_7 \times \mathfrak{su}_2$ \label{sec:E7SU2}}

In this section we turn to the F-theory realization of 6D vacua with hypermultiplets transforming in the
$(\mathbf{56},\mathbf{2})$ representation of $\mathfrak{e}_7 \times \mathfrak{su}_2$. As we have
already remarked in section \ref{sec:PRESCRIPTION}, the construction in F-theory is somewhat subtle, in
part because such bifundamentals appear to reside at singular points where the elliptic fiber is
no longer in Kodaira--Tate form.

On the other hand, we are in some sense guaranteed to find a sensible F-theory realization because there is a
well-known way to generate such bifundamental representations via compactifications of the $E_8 \times E_8$
heterotic string on the orbifold geometry $T^4 / \mathbb{Z}_2$. The full 6D theory has gauge algebra
$\mathfrak{e}_7 \times \mathfrak{su}_2 \times \mathfrak{e}_8$, a single tensor multiplet of charge $-12$, and hypermultiplets in the
representations
\begin{equation} \label{eq:perturbative_spectrum_E7xSU2}
(\mathbf{56},\mathbf{2},\mathbf{1}) \oplus (\mathbf{56}, \mathbf{1},\mathbf{1})^{\oplus 8} \oplus (\mathbf{1},\mathbf{2},\mathbf{1})^{\oplus 32} \oplus (\mathbf{1},\mathbf{1},\mathbf{1})^{\oplus 4}.
\end{equation}
Since the orbifold $T^4 / \mathbb{Z}_2$ can be viewed as an elliptically fibered K3 surface with tuned values of the K3 moduli, we expect
there to be a corresponding F-theory dual, and therefore, for an F-theory realization to exist. The resulting elliptically fibered
Calabi--Yau threefold was identified in \cite{Ludeling:2014oba}, but as we shall shortly review, it appears to contain singular collision points
which would ordinarily be identified with the presence of a 6D SCFT. However, by moving in the T-brane branch of moduli space, we can instead pass to a point where we retain a weakly coupled gauge theory description in the low energy effective field theory. This helps to clarify several aspects on the moduli space of these solutions, both in the F-theory and heterotic realizations of this model.

The rest of this section is organized as follows. First, we present some general considerations on the behavior of T-branes in F-theory backgrounds where the heterotic dual geometry is an orbifold. After this, we turn to the study of the moduli space of the
F-theory background which contains the $(\mathbf{56}, \mathbf{2})$, focussing in particular on the family of elliptically fibered
Calabi--Yau threefolds with $\mathbb{F}_{12}$ base \cite{Bershadsky:1996nh}. By tuning to a singular point in moduli space,
we reach a theory coupled to 6D SCFTs, as well as
its Higgs branch and tensor branch deformations. This setting is dual to heterotic
compactification on a K3 with 24 mobile instantons, of which the $T^4/\mathbb{Z}_2$ orbifold is a special limit \cite{Braun:2009wh}.

\subsection{T-Branes and Heterotic Orbifolds}

To frame the discussion to follow, we now briefly review some qualitative elements of our proposal. Recall that for heterotic strings compactified
on an elliptically fibered K3 surface with vector bundle given by the standard embedding (the tangent bundle $T_{\text{K3}}$),
the corresponding F-theory dual is described by a singular Weierstrass model with local presentation \cite{Morrison:1996na, Morrison:1996pp}:
\begin{equation}
y^2 = x^3 + g_{24}(s) t^5 + f_\text{K3} t^4 x + g_\text{K3} t^6 + ...
\end{equation}
with $f_\text{K3}$ and $g_\text{K3}$ the parameters of the minimal Weierstrass model for the dual heterotic K3 geometry.
To really specify the geometry, we also need to consider generic points in the complex structure moduli, wherein we work at a non-zero value
of the intermediate Jacobian. At the singular point in complex structure moduli, the associated T-brane deformation dissolves the small instantons
localized at the zeros of $g_{24}$ to finite size, leading to the heterotic tangent bundle model.
Already in this case, we can see that in spite of the singular presentation near
various $(4,6,12)$ points, we expect to have a weakly coupled description.

From this starting point of the standard embedding, we can ask what will happen as we further tune the moduli of the K3 surface so that it degenerates to the orbifold $T^4 / \mathbb{Z}_2$. There are two sorts of tunings we need to arrange to actually generate the standard heterotic orbifold. First, we seek a limit of an elliptically fibered K3 surface which has degenerated to this orbifold (see, e.g., \cite{Braun:2009wh}). Second, we must demand that the vector bundle used in the standard embedding also degenerates in a suitable fashion to reach this special point in moduli space. In the dual F-theory geometry, we expect to encounter some features such as T-branes, so we should not discard geometries where the multiplicity of vanishing for $(f,g,\Delta)$ is at least $(4,6,12)$. That being said, one way to understand such configurations is to first begin with a generic unfolding, and then perform further tunings in the discriminant locus. With this in mind, we shall first begin with a minimal Weierstrass model where the
$\mathfrak{e}_7$ is visible in the complex structure:
\begin{equation}
y^2 = x^3 + f_{20}(s) t^3 x + g_{24}(s) t^5 + f_\text{K3} t^4 x + g_\text{K3} t^6 + ...
\end{equation}
Further tuning in the parameters $f_{20}, g_{24}, f_\text{K3}, g_\text{K3}$ will then take us to the proposed heterotic
dual geometry. An important subtlety in this procedure is that because the heterotic dual has orbifold singularities, we must allow for the possibility that even though we are dealing with a T-brane deformation, the $f_{20}$ deformation parameters are actually ``visible'' in our presentation. Indeed, from a field theory perspective, the appearance of local $\mathbb{C}^{2} / \mathbb{Z}_2$ singularities in the heterotic geometry means that we are dealing with breaking patterns $\mathfrak{e}_{8} \times \mathfrak{su}_2 \rightarrow \mathfrak{e}_{7} \times (\mathfrak{su}_2)_{\text{diag}}$, where the diagonal $(\mathfrak{su}_{2})_{\text{diag}}$ is ``visible'' in the heterotic and F-theory geometries.

With these preliminaries dealt with, we now turn to the construction of an exotic bifundamental in our geometry.

\subsection{Generic Weierstrass Model with $\mathfrak{e}_7 \times \mathfrak{su}_2$}

To set up the notation, note that a Hirzebruch surface $\mathbb{F}_n$ is a $\mathbb{P}^1_\text{fiber}$-fibration over a $\mathbb{P}^1_\text{base}$.
Denoting the homogeneous coordinates along the fiber by $(u,v)$ and those along the base by $(s,t)$, their scaling relations and associated divisor classes are given by the following table:
\begin{align}\label{tab:scaling_relations_Fn}
	\begin{array}{c||c|c|c|c}
		& u & v & s & t \\ \hline\hline \rule{0pt}{2ex}
		[v] & 1 & 1 & 0 & 0 \\ \rule{0pt}{1.5ex}
		[s] & n & 0 & 1 & 1
	\end{array}
\end{align}
The intersection numbers are $[v]^2 = -n, \, [s]^2 = 0,  \, [v] \cdot [s] = 1$. The geometry also has self-intersection
$+n$ curves with divisor class $[u] = [v] + n [s]$. The anti-canonical bundle has class $\overline{K} = [u] +[v] + [s]+ [t] = 2[v] + (n+2)[s]$.
For a Calabi--Yau elliptic fibration over $\mathbb{F}_n$, the Weierstrass functions $f$ and $g$ have to have classes $4\overline{K}$ and $6\overline{K}$, respectively.

For $\mathbb{F}_{12}$, it is well known that any such elliptic fibration has to have a type II$^*$ fiber over $\{v\}$, which in F-theory corresponds to a ``non-higgsable'' $\mathfrak{e}_8$ gauge factor without any charged matter (see references \cite{Bershadsky:1996nh, Morrison:2012np}). To engineer the $\mathfrak{e}_7 \times \mathfrak{su}_2$ gauge factor, note that in the heterotic spectrum \eqref{eq:perturbative_spectrum_E7xSU2}, there is precisely one bifundamental hypermultiplet.
In an F-theory model with this matter content, anomaly cancellation  requires that the intersection number between 
the $\mathfrak{e}_7$ divisor $[\omega]$ and the $\mathfrak{su}_2$ divisor $[\sigma]$ is 12.
Furthermore, the spectrum also dictates that both divisors have genus 0 and self-intersection number 12.
Therefore, a natural choice is that $[\omega] = [\sigma] = [u]$.
Exploiting the reparametrization freedom along $\mathbb{P}^1_\text{fiber}$, we can put the
$\mathfrak{e}_7$ at $u = 0$. This leaves the $\mathfrak{su}_2$ divisor to be a generic +12 curve on $\sigma \equiv d_0\,u + p_{12}\,v =0$, where $d_0$ is a constant and $p_{12}$ is a degree 12 homogeneous polynomial in $(s,t)$.\footnote{The constant $d_0$ could also be absorbed into $p_{12}$, so that $\sigma = u + p_{12}\,v$.
However, we keep it here in order to have the same notation as in \cite{Ludeling:2014oba}.}

The two exceptional groups can be straightforwardly realized by the Weierstrass model:
\begin{align}
	f = u^3 \, v^4 \, \tilde{f} \, , \quad g = u^5 \, v^5 \, \tilde{g} \, , \quad \Delta = u^9\,v^{10}\,( \underbrace{4\,v^2\,\tilde{f}^3 + 27\,u\,\tilde{g}^2}_{\tilde{\Delta}}) \, .
\end{align}
The divisor class of the remaining free coefficients are
\begin{align}\label{eq:residual_classes_E7xSU2_Tate}
	\begin{split}
		[\tilde{f}] &= [u] + 8\,[s] = [v] + 20\,[s] \, ,\\
		[\tilde{g}] &= [u] + [v] + 12\,[s] = 2[u] = 2[v] + 24\,[s] \, .
	\end{split}
\end{align}
As discussed in \cite{Ludeling:2014oba}, this model can be viewed as having higgsed the $\mathfrak{su}_2$ completely with the doublets in \eqref{eq:perturbative_spectrum_E7xSU2}.
Importantly, the number of singlets required to cancel the gravitational anomaly agrees with the number of complex parameters in $\tilde{f}$ and $\tilde{g}$ modulo the reparametrization degrees of freedom of $\mathbb{F}_{12}$.

We now employ Tate's algorithm \cite{Bershadsky:1996nh, Katz:2011qp} to enhance the fiber singularity over $\{\sigma\}$ to I$_2$.
Because of the classes \eqref{eq:residual_classes_E7xSU2_Tate}, the most general form of $\tilde{f}$ and $\tilde{g}$ is
\begin{align}
	\begin{split}
		\tilde{f} &= \tilde{f}_{20} \, v + \tilde{f}_8\,\sigma \, , \\
		\tilde{g} &= \tilde{g}_{24}\,v^2 + \tilde{g}_{12}\,v\,\sigma + \tilde{g}_0 \, \sigma^2 \, \\
		\stackrel{u = \frac{\sigma - p\,v}{d_0}}{\Longrightarrow} \tilde{\Delta} = & \frac{v^5}{d_0} \, ( \underbrace{4\,d_0\,\tilde{f}_{20}^3 - 27\,\tilde{g}_{24}^2 \,p_{12} }_{\tilde\Delta_0}) + \frac{3\,v^4}{d_0} \, ( \underbrace{ 4\,d_0\,\tilde{f}_{20}^2\,\tilde{f}_8 + 9\,\tilde{g}_{24}\,(\tilde{g}_{24} - 2\tilde{g}_{12}\,p_{12})}_{\tilde{\Delta}_1} )\,\sigma + \CO(\sigma^2) \, .
	\end{split}
\end{align}
Note that $\tilde{g}_0$ is not really an independent free parameter, because it can be absorbed in $\sigma$ by the coefficient $d_0$.
To obtain an $\mathfrak{su}_2$ on $\sigma =0$, we have to tune $\tilde{\Delta}_{0} = \tilde{\Delta}_{1} = 0$, keeping in mind that these are polynomials over $\bbP^1_\text{base}$ and thus elements of a unique factorization domain (UFD).
Explicitly, requiring $\tilde\Delta_0 =4\,d_0\,\tilde{f}_{20}^3 - 27\,\tilde{g}_{24}^2 \,p_{12}=0$ means that prime factors of $\tilde{f}_{20}$ have to contain those of $p_{12}$.
However, because of the cube, $\tilde{g}_{24}$ must then also contain a factor of $p_{12}$, i.e.,
\begin{align}\label{eq:tuning_su2_step_1}
	\tilde{f}_{20} = c_8\,p_{12} \, , \quad \tilde{g}_{24} = c_{12}\, p_{12} \quad \Longrightarrow \quad \tilde{\Delta}_0 = p_{12}^3 \, (4\,d_0\,c_8^3 - 27\,c_{12}^2) \, .
\end{align}
To set this to zero, we therefore need $c_8 \sim c_4^2$ and $c_{12} \sim c_4^3$ for an arbitrary degree 4 polynomial $c_4$.
The coefficients turn out to be such that
\begin{align}
	\tilde{f}_{20} = \frac{1}{48} \, p_{12}\,c_4^2 \, , \quad \tilde{g}_{24} = \frac{\sqrt{d_0}}{864} \, p_{12} \, c_4^3 \, ,
\end{align}
which, as can be verified, sets $\tilde{\Delta}_0 = 0$.
Substituting into $\tilde{\Delta}_1$ yields
\begin{align}
	\tilde\Delta_1 \sim \sqrt{d_0}\,(c_4^2 + 144\,\tilde{f}_8)\,c_4  - 1728\,\tilde{g}_{12}  \stackrel{!}{=} 0 \, ,
\end{align}
which requires $\tilde{g}_{12} = \frac{\sqrt{d_0}}{1728} \,c_4\,(c_4^2 + 144\,\tilde{f}_8 )$.
Making the redefinitions $c_4 = 12\,d_4/\sqrt{d_0} $, $c_8 = d_8/d_0$, $g_0 = 1/d_0$ (recall that $g_0$ was not independent of $d_0$), the result of the above tuning is the Weierstrass model
\begin{align}\label{eq:generic_weierstrass_E7xSU2_p12}
	\begin{split}
		f & = \frac{u^3\,v^4}{d_0} \, \left( d_0\,d_8\,u + (3 \, d_4^2 + d_8)\, p_{12} \, v \right) \, , \\
		g & = \frac{u^5\,v^5}{d_0} \left( d_0^2\,u^2 + d_0\,(d_4^3 + d_4\,d_8 + 2\,p_{12}) \, u\,v + (3\,d_4^3 + d_4\,d_8 + p_{12}) \, p_{12} \, v^2 \right) , \\
		\Delta & = \frac{u^9 \, v^{10}}{d_0^3} \, (d_0 \, u + p_{12}\,v)^2 \, \underbrace{\left( 27\,d_0^3 \, u^3 + (...)\,u^2\,v + (...)\,u\,v^2 + 4 \, (3\,d_4^2 + d_8)^3\,p_{12}\,v^3 \right)}_{\Delta_{\text{res}}} \, .
	\end{split}
\end{align}
The unspecified parameters are in the polynomials of the indicated degrees, $d_0, d_4, d_8, p_{12}$, which yields a total of 28.
Of these, 3 can be removed by the $\text{SL}(2,\mathbb{C})$ reparametrization freedom on $\bbP^1_\text{base}$.
By insisting on the $\mathfrak{e}_7$ and $\mathfrak{e}_8$ divisors be on $u=0$ and $v=0$, there is only the rescaling freedom left on the fiber $\bbP^1$, which removes another free parameter.
Thus, there are $28 - 4 = 24$ independent deformation parameters preserving the $\mathfrak{e}_7 \times \mathfrak{su}_2$ gauge group, i.e., there are 24 neutral hypermultiplets.

Charged hypermultiplets are localized at intersections of discriminant components. At the intersection $u = p_{12} = 0$ of the $\mathfrak{su}_2$ and $\mathfrak{e}_7$ divisors, the vanishing orders of $(f,g,\Delta) \sim (4,6,12)$ indicate non-minimal singularities and will require further discussion.
To analyze other matter loci, we take a closer look at the I$_1$ locus over the residual discriminant component $\Delta_{\text{res}} =0$.
First, the only enhancement points on the $\mathfrak{e}_7$ divisor other than the non-minimal points are at $3d_4^2 + d_8=0$.
Importantly, the coefficient of $u\,v^2$ does not factor this polynomial, which also implies that the vanishing orders of the Weierstrass functions \eqref{eq:generic_weierstrass_E7xSU2_p12} at $u =0 = 3\,d_4^2 + d_8$ are $(f,g,\Delta) \sim (4,5,10)$, indicating the presence of $\frac{1}{2} \, {\bf 56}$'s.
Since $3d_4^2 + d_8$ has degree 8, there are therefore 8 half-hypermultiplets of $({\bf 56, 1})$ matter.

For the enhancement points on the $\mathfrak{su}_2$ divisor $\sigma = d_0 u +p_{12} v$, we rewrite the residual discriminant as
\begin{align}
	\Delta_{\text{res}} =  27 \, \sigma^3 + (...)\,\sigma^2 \, v + (...)\,\sigma\,v^2 + Q_{16}\,p_{12}\,d_4^2 \, v^3  \, ,
\end{align}
where neither the ellipses in parentheses nor the degree 16 polynomial $Q_{16}$ contain any overall factor of $d_4$ or $\sigma$.
Besides the non-minimal points at $p_{12}=0$, there are two other types of enhancements.
By simply checking the vanishing orders of the Weierstrass functions, we identify enhancements to I$_3$ resp.~type III at $Q_{16} = 0$ resp.~$d_4= 0$.
The enhancement to I$_3$ occur at normal crossings with the I$_1$ component at $\sigma = Q_{16} =0$ and thus lead to 16 hypermultiplets of $({\bf 1, 2})$.
On the other hand, the type III enhancement points with no matter are at order 2 tangency points $\sigma = d_4 =0$, where the residual discriminant goes like $\Delta_{\text{res}} \approx \sigma + d_4^2$.

Finally, the non-minimal points at $u = p_{12} = 0$ are transverse intersections between the $\mathfrak{e}_7$, $\mathfrak{su}_2$ divisors and the residual discriminant.
To resolve these, we have to blow-up the base with twelve exceptional divisors at each root of $p_{12} = u = 0$.
This introduces twelve $(-1)$-curves in the base, over which the elliptic fiber is easily checked to be smooth and thus no further matter states at their intersections with $\{u\}$ and $\{\sigma\}$.
Because all twelve blow-ups occur on the $\mathfrak{e}_7$ and $\mathfrak{su}_2$ divisors, each of their self-intersection number is lowered by 12, thus we end with self-intersection 0 curves in both cases.
This is consistent with anomaly cancellation, which requires precisely 4, resp.~16 fundamental hypermultiplets.
The twelve blow-ups further introduce twelve tensor multiplets, which contribute to the gravitational anomaly.
Taking into account the spectator $\mathfrak{e}_8$, we thus have:
\begin{align}\label{eq:grav_anomaly_E7-tate}
	\begin{split}
		T & = 13 \, , \\
		V & = 248 + 133 + 3 = 384 \, , \\
		H_{\text{charged}} & = 4 \cdot 56 + 16 \cdot 2 = 256 \, , \\
		H_{\text{neutral}} & = 24 \\
		\Longrightarrow \quad H - V +29\,T & = 273 \, .
	\end{split}
\end{align}

\subsection{Orbifold Conditions from Heterotic/F-theory Duality}

So far, we have not paid attention to the heterotic dual of this F-theory model on $\mathbb{F}_{12}$.
By heterotic/F-theory duality, the heterotic compactification space is a K3 surface inside the F-theory threefold that is elliptically fibered over $\bbP^1_\text{base}$.
To reach the $T^4 / \mathbb{Z}_2$ orbifold point in moduli space, the heterotic K3 first has to be tuned such that its elliptic fibration has four I$_0^*$ fibers.
Each such fiber contributes four volume moduli, which together with the volume of the base and the generic fiber yield 18 K\"ahler parameters, leaving two complex structure parameters inside $h^{1,1}(\text{K3})=20$.
Of these, 16 of the K\"ahler moduli have to be fixed to shrink all but the middle component of the I$_0^*$ fibers to zero size.
The four remaining free parameters (two complex and two K\"ahler) form the four singlets in the heterotic spectrum \eqref{eq:perturbative_spectrum_E7xSU2}.

As pointed out in \cite{Ludeling:2014oba}, one should identify the heterotic K3 inside the F-theory model by expanding the F-theory Weierstrass functions $(f,g)$ in the variables $(u,v)$,
\begin{align}
	f = \sum_{k \geq 0} f_{56-12\,k} \, u^k \, v^{8-k} \, , \quad g = \sum_{k \geq 0} g_{84-12\,k} \,u^k\, v^{12-k} \, ,
\end{align}
where the coefficients are homogeneous polynomials over $\mathbb{F}_{12}$'s base $\bbP^1$ of degree given by the subscript.
The elliptic fibration structure of the K3 is then described by the Weierstrass model $(f_\text{K3}, g_\text{K3}) = (f_8 , g_{12})$.
Furthermore, the 24 mobile instantons of the heterotic theory are localized in the fibers of the K3 over the zeroes of $g_{24}$.
For the heterotic K3 to have I$_0^*$ fibers at four points, which we can without loss of generality assume to be the roots of a non-degenerate degree four polynomial $p_4$ over $\bbP^1_\text{base}$, we need to tune $f_8 \sim p_4^2$ and $g_{12} \sim p_4^3$.
Additionally, the mobile instantons have to be trapped at the fixed points in the orbifold limit.
These fixed points correspond to the singularities of the K3 which are in the I$_0^*$ fibers, hence we also need $g_{24} \sim p_4^6$.

Applying these criteria to the generic Weierstrass model \eqref{eq:generic_weierstrass_E7xSU2_p12} with $\mathfrak{e}_7 \times \mathfrak{su}_2$, we obtain the conditions
\begin{align}
	\begin{split}
		\tilde{f} = d_0\,d_8 \sim p_4^2 \, , \quad \tilde{g} = d_0 (d_4^3 + d_4\,d_8 + 2\,p_{12}) \sim p_4^3 \, , \quad g_{24} = (3 d_4^3 + d_4 \, d_8 + p_{12}) \, p_{12} \sim p_4^6 \, ,
	\end{split}
\end{align}
The minimal (i.e., removing the least number of free parameters) tuning necessary is
\begin{align}\label{eq:final_tuning_E7xSU2}
	d_4 =\beta \, p_4 \, , \quad d_8 =  \alpha\, p_4^2 \, , \quad p_{12} =  p_4^3 \, ,
\end{align}
for some complex numbers $\alpha,\beta$.
The resulting elliptic fibration describes the F-theory geometry 
dual to the $T^4 /\mathbb{Z}_2$ orbifold \cite{Ludeling:2014oba}:
\begin{align}\label{eq:E7xSU2_LudelingRuehle}
	\begin{split}
		f &= \frac{p_4^2}{d_0} \, u^3 \, v^4  \, (d_0 \,u + (\alpha + 3\,\beta^2)\,p_4^3 \,v) \, , \\
		g &= \frac{1}{d_0} \, u^5 \, v^5 \, (d_0^2\,u^2 + d_0\,(2+\alpha\,\beta+\beta^3)\,p_4^3\,u\,v +  (1+\alpha\,\beta+ 3\,\beta^3)\,p_4^6\,v^2) \, ,\\
		\Delta &= \frac{1}{d_0^3} \, \underbrace{u^9}_{E_7} \, \underbrace{v^{10}}_{E_8} \, (\underbrace{d_0\,u + p_4^3\,v}_{SU(2)})^2 \, (\underbrace{\lambda_1 \, u^3 + \lambda_2\,p_4^3 \,u^2\,v + \lambda_3 \, p_4^6 \, u \,v^2 + \lambda_4\,p_4^9\,v^3}_{\Delta_\text{res}}) \,  \\
		\text{with} \qquad & \lambda_1 = 27\,d_0^3  \, , \quad \lambda_2 = 54\,d_0^2 \, (1 + \alpha\,\beta + \beta^3) \, , \quad \lambda_4 = 4\,(\alpha + 3\,\beta^2)^3 \, , \\
		& \lambda_3 = d_0\,(27 + 4\,\alpha^3 + 54\,\alpha\,\beta + 27 \, \alpha^2 \, \beta^2 + 162\,\beta^3 + 54\,\alpha \, \beta^4 + 27 \, \beta^6) \, .
	\end{split}
\end{align}
This Weierstrass model has eight free parameters in terms of $p_4, d_0, \alpha, \beta$.
Subtracting the $3+1$ reparametrization choices left on $\mathbb{F}_{12}$, it leaves precisely 4 singlets, as required by the perturbative heterotic spectrum \eqref{eq:perturbative_spectrum_E7xSU2}.

By inspection of the minimal Weierstrass model, we see various non-minimal points with order of vanishing for $(f,g,\Delta)$ at least 
$(4,6,12)$. In this respect, our analysis supplements the considerations of \cite{Ludeling:2014oba} 
because we have identified the appearance of T-brane deformations, which move 
us away from the singular point in moduli space. At such a point, blowing up the base 
is not physically allowed. This is in accord with the considerations of reference \cite{Ludeling:2014oba}. 
Of course, we can also move back to the singular point of the Higgs branch moduli space,
and then move onto the tensor branch. We turn to an analysis of this next.

\subsection{The Strongly Coupled Sector of the Global Model}\label{sec:discussion_E7xSU2}

The above tuning process was solely justified through the duality to heterotic theory.
In the following, we will discuss the model from a purely F-theory perspective and elucidate the connection to strongly coupled physics.

One key feature of the tuning process is that it changes the type of intersection between the $\mathfrak{e}_7$ divisor with the other discriminant components from transverse to tangential with order three.
From an F-theory perspective, this seems to be a necessary condition for the appearance of matter in the ${\bf 56}$ of $\mathfrak{e}_7$.
Indeed, the standard Weierstrass model with $\mathfrak{e}_7$ on a divisor $\{w\}$, $y^2 = x^3 + w^3\,\tilde{f}\,x + w^5 \, \tilde{g}$, has a residual discriminant of the form $\tilde{\Delta} = 27\tilde{g}^2 w + 4\tilde{f}^3$.
For generic $\tilde{g}$, the residual discriminant at $w = \tilde{f} = 0$ is locally described by the curve $w \sim \tilde{f}^3$, indicating that the 7-brane along $\tilde{\Delta}=0$ is intersecting the $\mathfrak{e}_7$ stack at $w=0$ with an order three tangency.
In the generic case, the singularity of the fiber enhances to $\mathfrak{e}_8$,
which according to the Katz--Vafa rules \cite{Bershadsky:1996nu, Katz:1996xe},
gives rise to a hypermultiplet of $\frac{1}{2}\,{\bf 56}$.

From this perspective, it is expected that the ``generic'' Weierstrass model \eqref{eq:generic_weierstrass_E7xSU2_p12} with $\mathfrak{e}_7 \times \mathfrak{su}_2$ gauge group cannot support matter in the $\mathbf{56}$ at the intersection between the divisors.
In that case, the residual discriminant factors into two components.
While the I$_1$ part still intersects the $\mathfrak{e}_7$ divisor $\{u\}$ tangentially at $u = 3\,d_4^2 + d_8=0$ , giving rise to ${\bf 56}$'s, the I$_2$ component along $\{d_0 u + p_{12} v\}$ now intersects $\{u\}$ transversely at $u = p_{12} = 0$.
Note that the global nature of the model forces the I$_1$ to also intersect $\{u\}$ transversely at this point, preserving the multiplicity three of the total intersection at each meeting point between the $\mathfrak{e}_7$ and the rest of the discriminant.
The resulting non-minimal singularity can only be resolved by blowing up the base.

However, the tangency argument alone is not enough to justify the tuning \eqref{eq:final_tuning_E7xSU2} that the heterotic dual requires.
This can be easily seen as it would have sufficed to set $p_{12} = p_4^3$ in \eqref{eq:generic_weierstrass_E7xSU2_p12}, which would have made the $\mathfrak{e}_7$ divisor intersect the I$_2$ and I$_1$ discriminant components tangential with order three.
So how can we make sense of the additional tunings $d_4 \sim p_4 , d_8 \sim p_4^2$ that are required to reach the orbifold limit in F-theory?
To shed some light on this issue, it proves to be instructive to again resolve the non-minimal points by blow-up, i.e., we go onto the tensor branch of the theory.
It is straightforward to check that the partial tuning $p_{12} = p_4^3$
does not change the gauge sector on the tensor branch:\footnote{The charge lattice of strings does change, however.}
Though we have grouped the 12 non-minimal points into four groups of three, the vanishing order of $(f,g,\Delta)$ is still $(4,6,12)$.
These points are still resolved with 12 exceptional curves, now three at each point, over all of which the elliptic fiber remains smooth.
Therefore, the numbers of tensors, hypers and vectors remain as in \eqref{eq:grav_anomaly_E7-tate}
in order to cancel the gravitational anomaly, i.e., we have not really lost the singlets by going from $p_{12}$ to $p_4^3$.

\begin{figure}[p]
	\centering
	\includegraphics[width=.9\hsize]{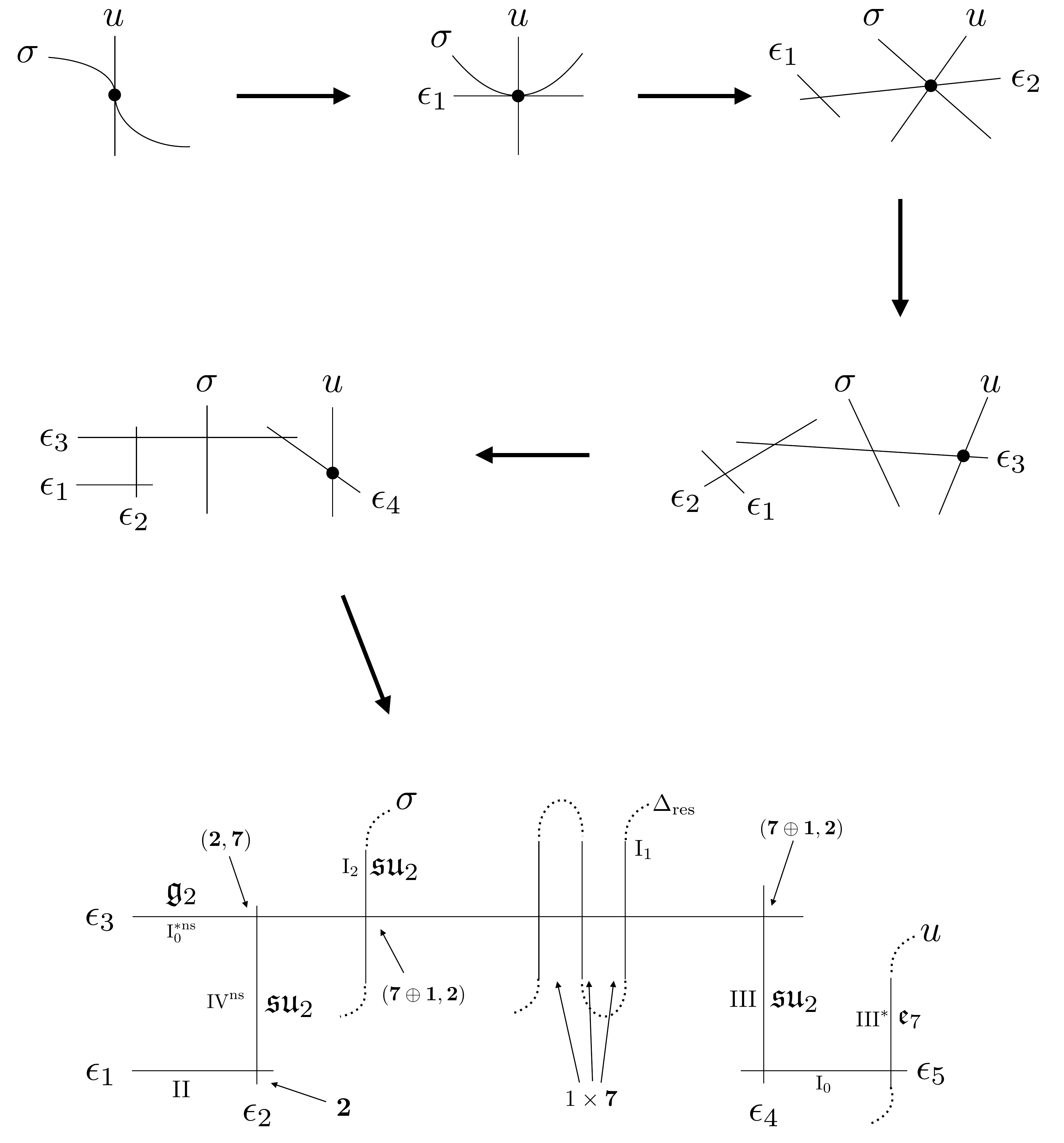}
	\caption{Blow-up of the base at one of the four points $u=p_4=0$, where the $\mathfrak{e}_7$ divisor $\{u\}$ and the $\mathfrak{su}_2$ divisor $\{\sigma \equiv d_0\,u + p_4^3 \,v\}$ intersect.
	It requires five $\bbP^1$'s $\{\epsilon_i\}$, carrying the indicated Kodaira fibers and associated gauge algebras, to resolve all non-minimal points (black dots).
	After the blow-ups the curve $\{u\}$ has self-intersection $-8$, and $\{\sigma\}$ has $0$.
	All of the exceptional curves have self-intersection $-2$, except for $\{\epsilon_5\}$ which has $-1$.
	All displayed matter multiplets are half-hypermultiplets, except the ${\bf 7}$ of $\mathfrak{g}_2$ on $\{\epsilon_3\}$, which is a full hypermultiplet spread across the three intersection points with the residual discriminant.
	}
	\label{fig:blow-ups_E7xSUs}
\end{figure}

The situation changes considerably when we include also the other two tuning conditions.
By setting $d_4 \sim p_4 , d_8 \sim p_4^2$, all the codimension two enhancement loci --- those supporting the $\bf (56,1)$ and $\bf (1,2)$ matter states as well as the type III fibers on the $\mathfrak{su}_2$ locus --- are collapsed onto the non-minimal point at $u = p_4 =0$.
The vanishing orders are now enhanced to $(f,g,\Delta) \sim (6,7,14)$.
By again blowing up these points in the base, we now observe that the tensor branch theory has changed drastically:
Now we need five blow-ups at each point to resolve the singularity, and these curves are decorated with singular Kodaira fibers over them.
The situation is summarized in figure \ref{fig:blow-ups_E7xSUs}.
The result of this blow-up process is a new gauge sector at each previously non-minimal point, consisting of the gauge factors $\mathfrak{su}_2 \times \mathfrak{g}_2 \times \mathfrak{su}_2$ with matter $\frac{1}{2} [ ({\bf 2,1,1}) \oplus ({\bf 2,7,1}) \oplus ({\bf 1, 7 \oplus 1, 2}) ] \oplus ({\bf 1,7,1}) $.
In addition, the $\mathfrak{su}_2$ divisor $\{\sigma\}$ passes through all four $\mathfrak{g}_2$ divisors, with each intersection contributing a $\frac{1}{2} ({\bf 7 \oplus 1, 2})$ hypermultiplet.
One can check that the gauge anomalies of each of the blow-up sectors is cancelled with this spectrum (all gauge groups introduced by blow-ups are on $-2$ curves).
Through the blow-up process, the self-intersection of the $\mathfrak{e}_7$ divisor has decreased to $12 - 4 \cdot 5 = -8$, which is required by the anomaly conditions for no charged matter.
The self-intersection of the I$_2$ divisor on $\sigma$ changes to $12 - 4\cdot 3 = 0$ --- the first three blow-ups are enough to remove any non-minimal points on it (see figure \ref{fig:blow-ups_E7xSUs}) --- which is also consistent with having in total $4 \times \frac{1}{2} \times 8 =16$ fundamental hypermultiplets.
The full spectrum on the tensor branch is thus
\begin{align}\label{eq:anomaly_E7_tuned_tensor_branch}
	\begin{split}
		H_\text{charged} & = 4 \times \# \left( \frac{1}{2} \,  [ ({\bf 2,1,1}) \oplus ({\bf 2,7,1}) \oplus ({\bf 1, 7 \oplus 1, 2}) ] \oplus ({\bf 1,7,1}) \oplus \frac{1}{2} ({\bf 7 \oplus 1, 2}) \right) \\
		& = 124 \, , \\
		V & = \dim \left( \mathfrak{e}_7 \times \mathfrak{su}_2 \times  \mathfrak{e}_8 \times ( \mathfrak{su}_2 \times \mathfrak{g}_2 \times \mathfrak{su}_2 )^{\otimes 4} \right) = 464 \, , \\
		T & = 1+ 4\cdot 5 = 21 \, ,\\
		\Longrightarrow  \, 273 & = H_\text{charged} - V + 29\,T + 4 \, ,
	\end{split}
\end{align}
which confirms that there are only four singlets in the theory.

The strongly coupled sector generated in this way is also interesting in its
own right. Taking a decompactification limit for the curves supporting the
$\mathfrak{e}_{7}$ and $\mathfrak{su}_2$ gauge algebra factors, we are left with the
tensor branch of a 6D\ SCFT (actually four copies) with tensor branch
description (with notation as in reference \cite{Heckman:2013pva, DelZotto:2014hpa, Heckman:2015bfa}
as well as the review \cite{Heckman:2018jxk}):
\begin{equation}\label{eq:strongly_coupled_sector_E7xSU2}
	[E_{7}] , 1 , \overset{\mathfrak{su}_{2}}{2}, \overset{\mathfrak{g}_{2}}{\underset{[Sp_{2}]}{2}} , \overset{\mathfrak{su}_{2}}{2} , 2
\end{equation}
In this SCFT there is an emergent $Sp_2 \simeq Spin_5$ flavor symmetry. Indeed, we have a
half hypermultiplet in the representation $(\mathbf{7},\mathbf{4})$ of $\mathfrak{g}_{2}%
\times\mathfrak{sp}_{2}$. The $\mathfrak{sp}_{2}$ flavor symmetry has a
maximal $(\mathfrak{su}_{2})_{L}\times(\mathfrak{su}_{2})_{R}$ subalgebra, one
factor of which is identified with the gauge algebra of the compact model.
From this perspective, our discussion also provides additional examples of 6D SCFTs coupled to
gravity. See e.g. \cite{DelZotto:2014fia, Anderson:2015cqy, Anderson:2018heq}
for some earlier examples. Note also that in the limit where
we decouple gravity (and the $E_7$ and $Sp_2$ are flavor symmetries) and proceed back to the SCFT point,
the $(\mathbf{56}, \mathbf{2})$ is a decoupled free hypermultiplet which is delocalized across the four collision
points of the $\mathfrak{e}_7$ and $\mathfrak{su}_2$ loci.
Such scenarios, where single hypermultiplets are spread across several enhancement points,  are not uncommon; in fact we have just seen above a similar fate for the $\bf 7$-representation of $\mathfrak{g}_2$ in the strongly coupled sector over three points.
However, such a delocalization of matter states is usually associated with matter charged under non-simply laced algebras, where we have the well-understood effect of Tate-monodromy.
For the gauge algebra $\mathfrak{e}_7 \times \mathfrak{su}_2$, there
is no known geometric mechanism which spreads one hypermultiplet across four codimension two loci.
This further emphasizes the importance of the global construction, as well as the non-local nature of the relevant T-brane deformation, which emerges only when we gauge the diagonal subalgebra in the flavor symmetry of four strongly coupled sectors of type \eqref{eq:strongly_coupled_sector_E7xSU2}.

With this discussion in hand, we can summarize the significance of
the tuning \eqref{eq:final_tuning_E7xSU2} as enhancing
the strongly coupled sector, which is always accessible via the tensor branch.
However, this highly tuned and singular F-theory model is
anomaly equivalent to the heterotic orbifold with a perturbative spectrum \eqref{eq:perturbative_spectrum_E7xSU2}.
This is a strong indication that at the origin of the tensor branch of the tuned model \eqref{eq:E7xSU2_LudelingRuehle} the strongly coupled sector has a Higgs branch, on which the heterotic orbifold model with the perturbative spectrum \eqref{eq:perturbative_spectrum_E7xSU2} sits.
In reference \cite{Ludeling:2014oba}, a matching of the spectrum was provided by performing various complex structure deformations, and then tuning back to the
singular point in complex structure moduli. From this perspective, we can instead consider deformations involving the intermediate Jacobian. In this case, the general description provided in \cite{Anderson:2013rka} (see also \cite{Anderson:2017rpr}) indicates that in the limit where the complex structure moduli are tuned to a singular value, we can remain in a weakly coupled regime due to the presence of a T-brane deformation.

\subsection{Adding More Bifundamentals}

The presentation given thus far has relied on the existence of a globally
complete heterotic dual. In this case, we observe that we can manage to get a
single bifundamental, which is delocalized across four points of the geometry.
It is natural to ask whether we can modify this construction to produce
examples with more than one bifundamental.

Considering that our criterion is primarily local in nature, we can in principle
relax any constraints on the
self-intersection numbers of the curves supporting the $\mathfrak{e}_{7}$ and
$\mathfrak{su}_2$ locus to produce local
geometries with these gauge symmetries and a bifundamental between them. Of course,
it will be more challenging to construct a fully compact model, and given the fact
that the curves supporting our gauge divisors have positive
self-intersection number, there is a sense in which 6D supergravity must
always be reintroduced.

Focussing on the purely gauge theoretic aspects of 6D anomaly cancellation,
the relevant constraints are (see e.g. \cite{Bershadsky:1996nh, Sadov:1996zm}):
\begin{align}
\begin{split}
\left(  \frac{D_{\mathfrak{e}_{7}}\cdot D_{\mathfrak{e}_{7}}}{2}+4\right)   &
=N_{(\mathbf{56},\mathbf{1})}+2\times N_{(\mathbf{56},\mathbf{2})}\\
\left(  6D_{\mathfrak{su(}2)}\cdot D_{\mathfrak{su(}2)}+16\right)   &
=N_{(\mathbf{1},\mathbf{2})}+56\times N_{(\mathbf{56},\mathbf{2})}\\
D_{\mathfrak{e}_{7}}\cdot D_{\mathfrak{su(}2)}  &  =12\times N_{(\mathbf{56}%
,\mathbf{2})},
\end{split}
\end{align}
where here, $D_{\mathfrak{g}}$ denotes the divisor class of a curve wrapped by
a seven-brane with gauge algebra $\mathfrak{g}$, and the $N$'s indicate the
multiplicity of various representations.

Increasing the number of bifundamentals necessarily raises the
self-intersection number relative to the case of a single bifundamental. To
see why, consider the extreme example where all matter comes from a
bifundamental.\ This can be satisfied if the divisors supporting gauge
algebras intersect as:%
\begin{align}
\begin{split}
	N_{(\mathbf{56},\mathbf{2})}  &  =2 \, , \quad N_{(\mathbf{56},\mathbf{1})}=0\, , \quad N_{(\mathbf{1},\mathbf{2})}=0 \, , \\
	D_{\mathfrak{e}_{7}}\cdot D_{\mathfrak{e}_{7}}  &  =0 \, , \quad D_{\mathfrak{su(}2)}\cdot D_{\mathfrak{su(}2)}=16 \, , \quad D_{\mathfrak{e}_{7}}\cdot D_{\mathfrak{su(}2)}=24 \, .
\end{split}
\end{align}
At this level of analysis, there are many examples of this sort. Clearly,
gravitational anomalies will impose additional conditions, but if we permit
ourselves to consider arbitrary bases, it is a priori possible that all such
anomaly cancellation conditions can be satisfied. We leave a full analysis of
such examples for future work.

We can also descend to 4D vacua by compactifying on a matter curve, and
in such models the constraints on the number of exotic bifundamentals are likely
substantially weaker. The construction of a Calabi--Yau fourfold with the requisite
singularities will of course be highly tuned, and will exhibit singularities
along curves (see e.g. \cite{Morrison:2016nrt}), and possibly at various Yukawa points
as well \cite{Apruzzi:2018oge}. Restricting our attention to a purely field theory
problem decoupled from gravity, we note that compactification of a 6D hypermultiplet on a genus $g$
curve\ $\Sigma$ yields, in the absence of any fluxes, $h^{0}(K_{\Sigma}%
^{1/2})=g$ 4D $\mathcal{N}=1$ chiral multiplets in the bifundamental
representation $(\mathbf{56},\mathbf{2})$, and $h^1(K_{\Sigma}^{1/2}) = h^{0}(K_{\Sigma}^{1/2})=g$
4D $\mathcal{N}=1$ chiral multiplets in the conjugate representation. From
this perspective, it is natural to expect there to be many ways to generate 4D
vacua with a large number of exotic bifundamentals.

\section{Engineering the $(\mathbf{27} , \mathbf{3})$ of $\mathfrak{e}_6 \times \mathfrak{su}_3$ \label{sec:E6SU3}}

As a second example, we construct an F-theory model that has bifundamental matter under $\mathfrak{e}_6 \times \mathfrak{su}_3$.
For that, we again rely on heterotic/F-theory duality to have a reference model.
Indeed, there is a heterotic $\mathbb{Z}_3$-orbifold which contains the desired gauge symmetry and matter.
Labeled as model IIId in \cite{Honecker:2006qz}, the gauge algebra is $\mathfrak{e}_6 \times \mathfrak{su}_3 \times \mathfrak{e}_7 \times \mathfrak{u}(1)$, with the following spectrum:\footnote{Note that the $\mathfrak{u}(1)$ charge here is normalized such that the only charged singlet has charge 1.
Compared to \cite{Honecker:2006qz}, this amounts to dividing the charge there by 2.}
\begin{align}\label{eq:E6xSU3_spectrum_heterotic}
	\begin{split}
		& ({\bf 27, 3, 1})_0  \oplus ({\bf 27, 1, 1})_{\frac{1}{3}}^{\oplus 9} \oplus ({\bf 1, 3, 1})_{\frac{2}{3}}^{\oplus 9}
\oplus ({\bf 1, 3, 1})_{-\frac{1}{3}}^{\oplus 18} \\
		\oplus  \, &({\bf 1,1,56})_{\frac{1}{2}} \oplus ({\bf 1}, \mathbf{1}, \mathbf{1})_1 \oplus
({\bf 1}, \mathbf{1}, \mathbf{1})_0^{\oplus 2} \, .
	\end{split}
\end{align}
Cancelling the resulting gauge anomalies in F-theory requires the self- and mutual intersection number of both the $\mathfrak{e}_6$ and $\mathfrak{su}_3$ divisors to be 6, as well as the self-intersection number of the $\mathfrak{e}_7$ divisor to be $-6$.
Thus, the natural choice for the F-theory base is $\mathbb{F}_6$, with $\mathfrak{e}_7$ on the unique $-6$-curve $\{v\}$ and the $\mathfrak{e}_6$ and $\mathfrak{su}_3$ divisors on $+6$-curves $\{u\}$ and $\{\sigma \equiv u + p_6\}$, respectively, with $p_6$ a homogeneous degree six polynomial on the $\bbP^1_\text{base}$ of $\mathbb{F}_6$.

Following the extensive discussion in the $T^4/\mathbb{Z}_2$ case, we will use the two criteria developed in section \ref{sec:discussion_E7xSU2},
namely the correct order of tangential intersections and the enhancement of the tensor branch theory, to construct the F-theory dual of the $\mathbb{Z}_3$-orbifold.
However, due to the global nature of $\mathfrak{u}(1)$'s in F-theory, our local criteria cannot fully determine the necessary tuning.
Physically, the missing tuning conditions are reflected in a miscount of the singlets,
which can be especially delicate in an interplay with a $\mathfrak{u}(1)$.

In order to have a good starting point, we begin with a model over $\mathbb{F}_6$ having $\mathfrak{e}_6 \times \mathfrak{su}_3 \times \mathfrak{e}_6$ constructed via Tate's algorithm, as detailed in appendix \ref{app:Tate_E6xSU3}.
Here, the second $\mathfrak{e}_6$ is the non-higgsable algebra on the $-6$ curve $\{v\}$.
In order to enhance to $\mathfrak{e}_7$, as required for the heterotic orbifold, we simply set the parameter $g_0 = 0$.
The resulting model
\begin{align}\label{eq:E6xSU3_Tate}
	\begin{split}
		f  = \, & \left [f_2\, u^2 + (2\, f_2\, p_6 - 3\, c_2^4 + c_2\, c_6 )\, u\, v + p_6\, (f_2\, p_6 - c_2\, c_6)\, v^2 \right] \, u^3 \, v^3\, , \\
		g =\, & \left[ \left( 2\,\alpha\,c_6 + \alpha^2\,p_6 - 12\,f_2\,c_2^2 \right) u^3   \right. \\
       			&  + \left( 3\,\alpha^2 \,p_6^2 + 24\,c_2^2\,(c_2^4 - f_2\,p_6) + c_6 \, (c_6 + 12\,c_2^3 + 6\,\alpha\,p_6 ) \right)  u^2\,v \\
			  & + \left( 3\,\alpha^2\,p_6^2 + 2\,c_6 \, (6\,c_2^3 + c_6) + 6\,p_6 \, (\alpha\,c_6 - 2\,f_2\,c_2^2)  \right)  \,p_6\,u\,v^2   \\
      		&    \left. + p_6^2\,( c_6  + p_6\,\alpha)^2 \,v^3 \right] \, \frac{u^4 \, v^5}{12} \, , \\
		\Delta = \, & \frac{1}{16}\,u^8\,v^9\,(u+p_6\,v)^3 \, \Delta_{\text{res}} \, ,
	\end{split}
\end{align}
with
\begin{align}\label{eq:res_disc_E6xSU3_Tate}
\begin{split}
	\Delta_{\text{res}} & = 4\,f_2^3 u^4 + (...)\,u^3v + (...) \, u^2v^2 + (...)\, u\,v^3 + \frac{3\,p_6\,(\alpha \, p_6+ c_6)^4}{16}\, v^4 \\
	&= 4\,f_2^3\,\sigma^4 + (...)\,\sigma^3v + (...)\,\sigma^2 v^2 + (...)\,\sigma\,v^3 + p_6\,c_2^3\,Q_{18}\,v^4 \,  ,
\end{split}
\end{align}
where $\sigma = u + p_6 v$ and $Q_{18}$ a degree 18 polynomial.

The six non-minimal points at $u=p_6=0$ with vanishing orders $(f,g,\Delta) \sim (4,6,12)$ are a result of the transverse intersections between the $\mathfrak{e}_6$, the $\mathfrak{su}_3$ and the residual discriminant components.
They are resolved by a blow-up at each point with one exceptional curve, over which the elliptic fiber turns out to be smooth.
Thus these blow-ups do not introduce additional gauge algebras.
They do however separate $\{u\}$ and $\{\sigma\}$ completely and lower the self-intersection number of the $\mathfrak{e}_6$ and $\mathfrak{su}_3$ divisor both to $0$.
The gauge anomaly is then cancelled by the six ${\bf 27}$'s at $u = \alpha\,p_6 + c_6 = 0$ and $18 \times \bf 3$'s at $\sigma = Q_{18}=0$.
Furthermore, the $\mathfrak{e}_7$ gauge anomaly is cancelled by the two $\frac{1}{2} \bf 56$'s at $v = f_2 = 0$.
Finally, the gravitational anomaly requires 17 uncharged singlets, in agreement with the 21 free parameters in $f_2, c_2, c_6, p_6, \alpha$ modulo the four reparametrization degrees of freedom of the base $\mathbb{F}_6$.
In order to obtain the F-theory geometry dual to the heterotic orbifold, we have to specialize these parameters further.


The first step towards the perturbative heterotic spectrum \eqref{eq:E6xSU3_spectrum_heterotic} should now be to ensure the correct tangential intersection between the $\mathfrak{e}_6$ divisor and the other discriminant components.

\subsection{Tangency of Intersection and the Need for a Type IV Fiber}

Similar to the case of $\mathfrak{e}_7$, fundamental matter of $\mathfrak{e}_6$ generically appears at a tangential
intersection of the $\mathfrak{e}_6$ locus with another discriminant component.
While for $\mathfrak{e}_7$, the necessary order of tangency was three,
it is four for $\mathfrak{e}_6$ (as can be seen from a generic Weierstrass model with $E_6$).
However, for the I$_3$ locus on $\{u+p_6\,v\}$, there is no possible factorization
of $p_6$ such that one has an order four tangency at all points of intersection with $\{u\}$.

To resolve this issue, we propose that the $\mathfrak{su}_3$ should be realized with a type IV fiber.
First, the above naive picture applies to a single 7-brane hitting the $\mathfrak{e}_6$, such that the fundamental states can be thought of as massless strings ending on the $\mathfrak{e}_6$ branes and the single 7-brane.
Locally, in the I$_3$ version of $\mathfrak{su}_3$, the $\mathfrak{e}_6$ sees three mutually local 7-branes which happen to pass through the same point.
From the $\mathfrak{e}_6$ perspective, each of these branes would have to therefore meet at an order four tangency, which is impossible in the given setup.
However, the type IV version of $\mathfrak{su}_3$ is realized on four branes which are mutually non-local.
Thus, there is the possibility of multi-pronged strings that end on two of the four branes while still constituting a single state from the $\mathfrak{e}_6$ perspective.
Indeed, if one looks at a generic model with type IV $\mathfrak{su}_3$ on $\{w\}$,
\begin{align}\label{eq:generic_type_IV_SU3}
	f = w^2\,\tilde{f} \, , \quad g = w^2\,\tilde{g}^2 \, , \quad  \Delta = w^4 \, (\underbrace{4\,\tilde{f}^3\,w^2 + 27\,\tilde{g}^4}_{\tilde{\Delta}}) \, ,
\end{align}
one can see that the codimension two enhancement loci $\{w\} \cap \{\tilde{\Delta}\} = \{w = \tilde{g} =0\}$, which is known to host three hypermultiplets of triplets at each point, locally looks like two of the four branes on $\{w\}$ meeting the residual I$_1$ locus $\{\tilde{\Delta}\}$ at quadratic order.

On the $\mathbb{F}_6$ base, such a non-transversal meeting is indeed possible between $\mathfrak{e}_6$ and type IV $\mathfrak{su}_3$, both on $+6$-curves:
One simply has to set the $\mathfrak{su}_3$ locus to be $\{u + p_3^2\,v\}$, which
meets the $\mathfrak{e}_6$ locus on $\{u\}$ tangentially with order two.
In the above model \eqref{eq:E6xSU3_Tate}, setting $c_2 = 0$ enhances the singularity type over the $\mathfrak{su}_3$ locus to type IV, which then is guaranteed to intersect the $\mathfrak{e}_6$ at quadratic order with $p_6 = p_3^2$.
The Weierstrass model now takes the form
\begin{align}
		f & = f_2 \, u^3 \,v^3 \, (u+p_3^2 \,v)^2 \, , \notag \\
		g &= \frac{1}{12}\, u^4\,v^5\, (u+p_3^2\,v)^2 \, (\alpha \, (\alpha\,p_3^2 + 2\,c_6) \, u + (c_6 + \alpha \, p_3^2)^2\,v ) \, , \label{eq:E6xSU3_tangency_tuned} \\
		\Delta & = \frac{1}{16} \, u^8\,v^9 \, (u+p_3^2\,v)^4\, \left( 64\,f_2^3\,u\,(u+p_3^2\,v)^2 + 3\,v\,  (\alpha \, (\alpha\,p_3^2 + 2\,c_6) \, u + (c_6 + \alpha \, p_3^2)^2\,v ) ^2 \right) \, . \notag
\end{align}

Similar to the $\mathfrak{e}_7 \times \mathfrak{su}_2$ case, the tangency argument alone is not sufficient to obtain the correct orbifold dual model.
Again, we can convince ourselves that despite the tuning $p_6 = p_3^2$, the non-minimal points at $p_3 = u=0$, now with vanishing orders $(f,g,\Delta) \sim (5,6,12)$, are still resolved with six blow-ups (two at each point), with no additional gauge algebras or matter accompanying them (the intersection structure of these blow-ups do change).
In fact, counting the singlets naively one finds that there are three singlets missing to cancel the gravitational anomaly.
However, this is no surprise as we have set $c_2=0$ to obtain the type IV fiber, which does not change the charged spectrum on the tensor branch compared to the model with I$_3$ \eqref{eq:E6xSU3_Tate}.
Therefore, the three parameters of $c_2$ still correspond to massless singlets present in the tensor branch theory.

\subsection{Enhancing the Strongly Coupled Sector}

The next step of the tuning is now to enhance the gauge sector on the tensor branch, which becomes strongly coupled when we blow-down the exceptional curves.
To do that, we again collapse all codimension two enhancement loci on the $\mathfrak{e}_6$ and $\mathfrak{su}_3$ divisors on top of each other.
A quick look at \eqref{eq:E6xSU3_tangency_tuned} reveals that the $\mathfrak{e}_6$ intersects the residual discriminant at $c_6 + \alpha\,p_3^2 = u =0$, while the $\mathfrak{su}_3$ divisor meets it at $u + p_3^2\,v = 0 = c_6$.
To collapse both loci onto $u = p_3 =0$, we therefore have to tune $c_6 = \beta\,p_3^2$.
With some slight cosmetic redefinitions,\footnote{Define $\alpha \equiv \lambda - \sqrt{\lambda^2 - \mu}$ and $\beta \equiv \sqrt{\lambda^2 - \mu}$.} our F-theory geometry is now given by
\begin{align}\label{eq:E6xSU3IV_weierstrass_noU1}
	\begin{split}
		f & = f_2 \, u^3 \, v^3 \, (u+p_3^2\,v)^2 \, , \\
		g &= \frac{p_3^2}{12} \, u^4 \, v^5 \, (u+ p_3^2\,v)^2 \, ( \mu \, u + \lambda^2 \, p_3^2 \, v) \, , \\
		\Delta & = \frac{1}{16} \, u^8 \, v^9 \, (u+ p_3^2\,v)^4 \, \left( 64 \, f_2^3 \, u \, (u+p_3^2\,v)^2 + 3\,p_3^4 \, v \, (\mu \, u + \lambda^2\,p_3^2 \, v)^2  \right) \, .
	\end{split}
\end{align}

\begin{figure}[p]
	\centering
	\includegraphics[width=.9\hsize]{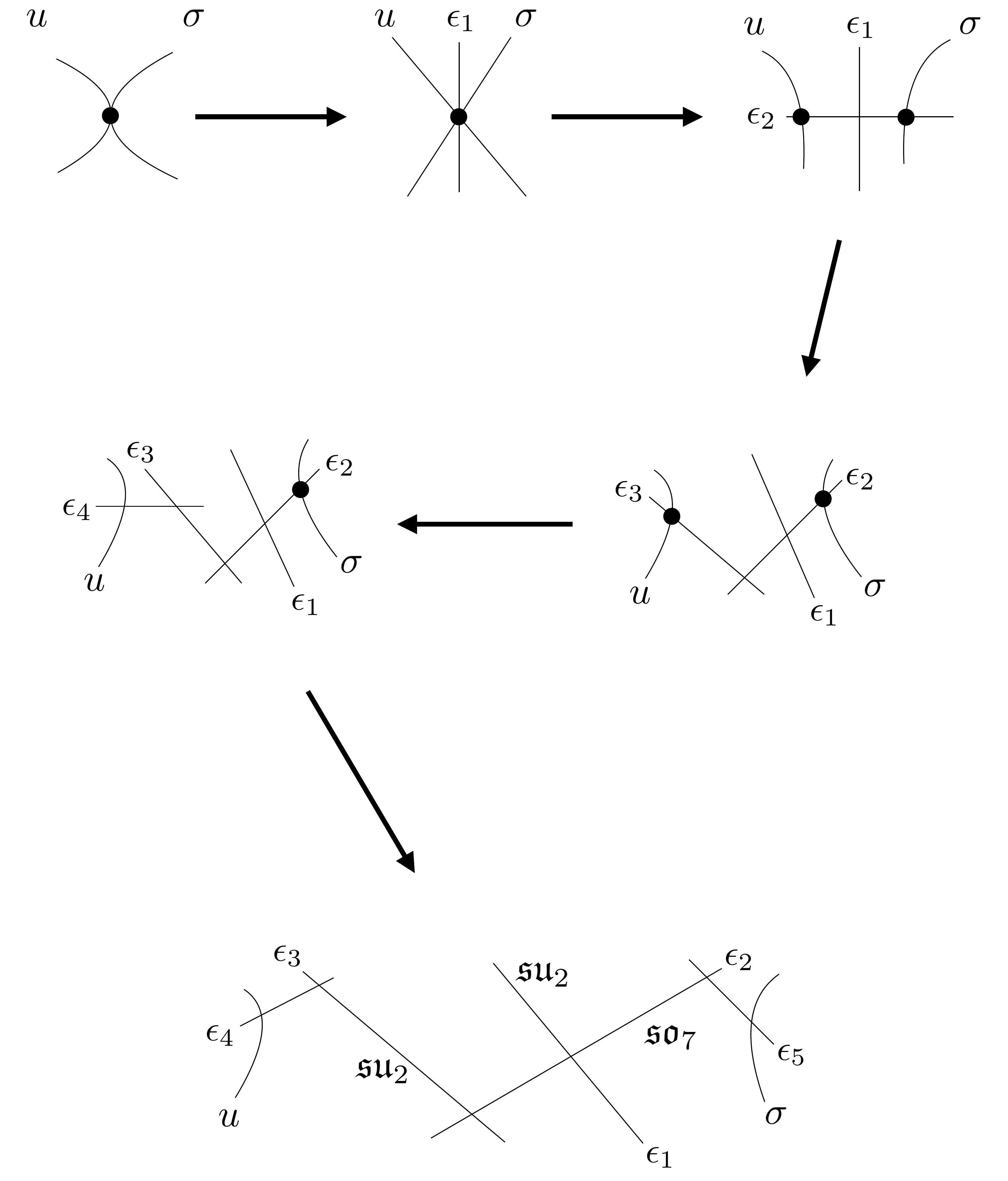}
	\caption{Blowing-up the base at the three points $u=p_3=0$ where the $\mathfrak{e}_6$ divisor $\{u\}$ and the $\mathfrak{su}_3$ divisor $\{\sigma \equiv u + p_3^2 \,v\}$ intersect.
	It requires five $\bbP^1$'s $\{\epsilon_i\}$ to resolve all non-minimal points (black dots).
	After the blow-ups the curve $\{u\}$ has self-intersection $-6$, and $\{\sigma\}$ has $-3$.
	The sequence of curves $\epsilon_1 - \epsilon_2 - \epsilon_3$ has self-intersections $(-2, -3, -2)$ and have the gauge algebras $\mathfrak{su}_2 - \mathfrak{so}_7 - \mathfrak{su}_2$.
	At the two intersection points we have $\frac{1}{2} ( {\bf 8,2})$, where ${\bf 8}$ is the spinor representation of $\mathfrak{so}_7$.
	}
	\label{fig:blow-ups_E6xSU3}
\end{figure}

Let us verify that this tuning indeed enhances the strongly coupled sector by blowing up the only non-minimal points at $u = p_3 =0$, which now have vanishing orders $(f,g,\Delta) \sim (5,9,15)$.
The result of this process is a tree of five rigid curves for each point, cf.~figure \ref{fig:blow-ups_E6xSU3}.
Inside there is a chain of $(-2) (-3) (-2)$ curves carrying singular fibers corresponding to the gauge algebras $\mathfrak{su}_2 - \mathfrak{so}_7 - \mathfrak{su}_2$.
Anomaly cancellation forces the matter content to be $\frac{1}{2} ( {\bf 2,8,1}) \oplus \frac{1}{2} ( {\bf 1,8,2})$.
The blow-ups lower the self-intersection numbers of the $\mathfrak{e}_6$ divisor to $-6$ and of the $\mathfrak{su}_3$ divisor to $-3$, so they are non-higgsable, i.e., have no matter \cite{Morrison:2012np}.
These blow-ups did not touch the $\mathfrak{e}_7$ loci, whose $2 \times \frac{1}{2} \bf 56$ are localized at $\{v = f_2 = 0\}$.
Finally, there are nine free parameters in $f_2, \, p_3, \, \lambda , \, \mu$, which after subtracting the four reparametrization degrees of freedom yield five uncharged singlets.
The complete spectrum is therefore:
\begin{align}
\begin{split}
	H_{{\bf 1}_0} & = 5 \, , \\
	H_{\text{charged}} & = \# \left( 3 \times \frac{1}{2} ( ( {\bf 2,8,1}) \oplus ({\bf 2,8,1}) ) + 2 \times \frac{1}{2} {\bf 56} \right)=  104 \, ,\\
	V & = 133 + 78 + 8 + 3 \, (3 + 21 + 3) = 300 \, ,\\
	T & = 1 + 3\cdot 5 = 16 \, ,\\
	\Longrightarrow  273 & = H_\text{charged} + H_{{\bf 1}_0}  - V + 29\,T\, .
\end{split}
\end{align}
It is reassuring to see that the singlets are counted correctly by our free
parameters, as indicated by the cancellation of the gravitational anomaly.

We can analyze further aspects of the strongly coupled sector by
decompactifying the curves supporting the $\mathfrak{e}_{6}$ and $\mathfrak{su}_3$ gauge groups.
Doing so, we are left with the tensor branch of a 6D\ SCFT (actually three
copies)\ with tensor branch description (with notation as in references
\cite{Heckman:2013pva, Heckman:2015bfa, Heckman:2018jxk}):
\begin{equation}
[E_{6}],1,\overset{\mathfrak{su}_{2}}{2},\overset{\mathfrak{so}%
_{7}}{\underset{[SO_{9}]}{\underset{1}{3}}},\overset{\mathfrak{su}_{2}}{2}
\end{equation}
Namely at the SCFT\ point, there is an emergent $SO(9)$ flavor symmetry.
Indeed, the $-1$ curve theory supports an E-string theory, and we can gauge
the $\mathfrak{so}_{7}$ factor of the $\mathfrak{so}_{7}\times\mathfrak{so}_{9}%
\subset\mathfrak{e}_{8}$ subalgebra. The $\mathfrak{su}_3$ gauge symmetry of the compact
model is obtained by gauging a subalgebra of this $SO(9)$ flavor symmetry.

\subsubsection*{Cross Check --- Dual Heterotic K3 as \boldmath{$\mathbb{Z}_3$-orbifold}}

At this point, it is worth comparing the above tuning, motivated solely from an F-theory perspective, with the heterotic K3 in the dual orbifold model. For the case at hand, the elliptic fibration of the K3 has to have three IV$^*$ fibers to be compatible with the $T^4/\mathbb{Z}_3$ orbifold (see e.g. \cite{Dasgupta:1996ij}). This means that the Weierstrass functions have to be of the form $(f_\text{K3}, g_\text{K3}) = (0, q_3^4)$, where the position of the singular fibers are the roots of some degree three polynomial $q_3$.
Furthermore, the heterotic dual of an F-theory model over $\mathbb{F}_6$ has 18 mobile instantons, which again have to be localized at the orbifold fixed points, i.e., inside the IV$^*$ fibers.

Naively, these conditions seems to be incompatible with \eqref{eq:E6xSU3_tangency_tuned}:
Identifying $f_\text{K3}$ as the coefficient of $u^4v^4$ in $f$ would then require setting $2f_2 p_3^2$ to 0.
However, by setting either of the two functions to zero would enhance the singularity type over either $\{v\}$ or $\{u\}$, none of which would yield the spectrum of the heterotic orbifold.

On the other hand, the curve $\{u\}$ is not special on the $\mathbb{F}_6$ base of F-theory.
Unlike the $-6$-curve $\{v\}$, $+6$-curves come in a family $\{s \equiv u + q_6 \,v = 0\}$ for any $q_6$.
Therefore, there is no obvious reason why one cannot identify the gluing K3 along any of these $+6$-curves $\{s\}$.
In fact, one might view this shift by $q_6$ as a consequence of the non-trivial shift inside both $E_8$'s on the heterotic orbifold side.
So we write the F-theory Weierstrass functions \eqref{eq:E6xSU3_tangency_tuned} as polynomials $f = \sum_k \, f'_{d_k} \, s^k \, v^{5-k}$, $g = \sum_k g'_{d_k} \, s^k \, v^{8-k}$.
Then, the heterotic K3 is the Weierstrass model with
\begin{align}
	\begin{split}
		f_{\text{K3}} & \equiv f'_8 = f_2\,(2\,p_3^2 - 5\,q_6) \, .
	\end{split}
\end{align}
The necessary condition $f_\text{K3} = 0$ can now be solve with $q_6 = 2/5\,p_3^2$.
It is reassuring to see that the resulting $g_\text{K3}$ as well as the polynomial $g'_{18}$ encoding the location of the 18 mobile instantons are compatible with the orbifold:
\begin{align}
	\begin{split}
		g_{\text{K3}} \equiv g'_{12} & = \frac{(5\,\lambda^2 - 4\,\mu)}{60} \, p_3^4\, , \\
		g'_{18} & = -\frac{10\,\lambda^2 + 11\,\mu}{300} \, p_3^6 \, ,
	\end{split}
\end{align}
i.e., the heterotic K3 has three IV$^*$ fibers at $p_3 =0$, at which the 18 instantons are localized.

\subsection{Engineering the $\mathfrak{u}(1)$ and Singlet Counting}

The only part of the gauge group that is still missing is the $\mathfrak{u}(1)$ factor.
To engineer it, the F-theory fibration has to have a non-trivial Mordell--Weil group, i.e.,
another rational section in addition to the zero section.
With the tuned model \eqref{eq:E6xSU3IV_weierstrass_noU1}, this turns out to be fairly easy.
By inspection of $g$, we see that it is almost a perfect square up to the factor $v^5 \, (\mu \, u + \lambda^2\,p_3^2 \,v)$.
However, this can be easily changed by setting $\mu = 0$, yielding
\begin{align}\label{eq:E6xSU3_withU1_wrong}
	\begin{split}
		f & = f_2 \, u^3 \, v^3 \, (u+p_3^2\,v)^2 \, , \\
		g &= \frac{\lambda^2}{12} \,p_3^4\, u^4 \, v^6 \, (u+ p_3^2\,v)^2 \,   \, , \\
		\Delta & = \frac{1}{16} \, u^8 \, v^9 \, (u+ p_3^2\,v)^4 \, \left( 64 \, f_2^3 \, u \, (u+p_3^2\,v)^2 + 3\,  \lambda^4 \,p_3^8 \, v^2   \right) \, .
	\end{split}
\end{align}
Then, the Weierstrass equation
\begin{align}
	y^2 = x^3 + f\,x + g
\end{align}
has a rational solution at $(x,y) = (0, \sqrt{g}) = (0 , \lambda\,p_3^2\,u^2 \, v^3 \, (u+p_3^2\,v)/\sqrt{12})$, which is clearly not the zero-section at infinity.

Setting $\mu=0$ makes the singularity over the locus $v=f_2=0$ non-minimal.
The blow-up introduces two tensors and no additional gauge group.
They do, however, change the self-intersection of the $\mathfrak{e}_7$
divisor to $-8$, which does not allow for any $\mathfrak{e}_7$ matter.
The change in the gravitational anomaly can also be computed:
\begin{align}
	\delta H - \delta V + 29 \delta T = (-56 - 1) - 1 + 2 \cdot 29 = 0 \, .
\end{align}

Unfortunately, this tuning is not enough to produce the heterotic dual, as reflected by the mismatch in the number of uncharged singlets:
By losing only one singlet compared to \eqref{eq:E6xSU3IV_weierstrass_noU1}, we still have four uncharged singlets, which exceeds the counting in the heterotic spectrum \eqref{eq:E6xSU3_spectrum_heterotic} by two.

To resolve this paradox, we need to pay more attention to the Higgs branch of the heterotic model.
Given the heterotic spectrum \eqref{eq:E6xSU3_spectrum_heterotic} with gauge algebra $\mathfrak{e}_6 \times \mathfrak{su}_3 \times \mathfrak{e}_7 \times \mathfrak{u}(1)$, one cannot higgs the $\mathfrak{u}(1)$ in a D-flat manner with just one charged singlet.
Consequently, any higgsing that breaks the $\mathfrak{u}(1)$ must proceed via vevs of the other matter multiplets, which necessarily also break parts of the non-abelian gauge group.
Conversely, this means that an honest complex structure tuning that corresponds to a chain of unhiggsing should have implemented the $\mathfrak{u}(1)$ before the appearance of all the non-abelian gauge factors.

One starting point could be a model with $\mathfrak{e}_6 \times \mathfrak{e}_7$ gauge symmetry only, given by the Weierstrass model
\begin{align}\label{eq:ft_gt_expansion_E6xSU3}
	\begin{split}
		f & = u^3\, v^3 \, \ft \, , \quad g = u^4\,v^5 \, \gt \, , \quad \text{with} \\
		\ft & = f_2 \, u^2 + f_8 \, u \,v + f_{14} \,v^2 \, , \\
		\gt & = g_6 \, u^3 + g_{12} \, u^2\,v + g_{18} \, u\,v^2 + \gamma_{12}^2 \, v^3  \\
		\Longrightarrow \, \Delta & = u^8\,v^9 \, ( 4 \,u^7 \, f_2^3  + 27\, v^7 \, \gamma_{12}^4 + u\,v\,(...) ) \, ,
	\end{split}
\end{align}
which certainly lies in the Higgs branch of the heterotic model (one can break the $\mathfrak{su}_3 \times \mathfrak{u}(1)$ with the charged triplets only).
This geometry was also be taken as the starting point of our tuning process, that led us to the above $\mathfrak{e}_6 \times \mathfrak{su}_3 \times \mathfrak{e}_7$ model \eqref{eq:E6xSU3IV_weierstrass_noU1} (see appendix \ref{app:Tate_E6xSU3}).
Thus one might hope that by first implementing the additional rational section, one could still incorporate the above tuning to yield the desired $\mathfrak{su}_3$ part.
Unfortunately, despite the many recent advances in the study of $\mathfrak{u}(1)$'s in F-theory,\footnote{Starting with \cite{Grimm:2010ez} and subsequently \cite{Morrison:2012ei, Mayrhofer:2012zy, Cvetic:2012xn} there have been many recent
developments. For a more complete set of references, see
the review article \cite{Weigand:2018rez}.} it is technically still very challenging to come up with a model that not only includes the $\mathfrak{u}(1)$, but also has no redundancies such that we can correctly count the uncharged singlets by the free parameters in the Weierstrass model.\footnote{For example, while the so-called Morrison--Park model \cite{Morrison:2012ei} guarantees a section by embedding the fiber into an $\text{Bl}_1 \bbP_{112}$ ambient space, the resulting Weierstrass model has some redundancies in terms of the coefficients of the $\text{Bl}_1 \bbP_{112}$-hypersurface, which have to be identified for a correct counting of uncharged singlets.}

However, there is one key feature of the $\mathfrak{u}(1)$ which is easy to implement and sufficient to identify one necessary complex structure tuning.
Starting from \eqref{eq:ft_gt_expansion_E6xSU3} and enhancing the gauge group to $\mathfrak{e}_6 \times \mathfrak{e}_7 \times \mathfrak{u}(1)$, the two $\frac{1}{2} \bf 56$ hypermultiplets have to merge into a single hypermultiplet charged under the $\mathfrak{u}(1)$.
Since such a spectrum is perfectly compatible with minimal Kodaira fiber types, standard F-theory dictates that a necessary geometric condition for the existence of such a $\mathfrak{u}(1)$ is the factorization $f_2 \rightarrow f_1^2$, which combines the two loci of the $\frac{1}{2} \bf 56$'s in the $\mathfrak{e}_6 \times \mathfrak{e}_7$ model \eqref{eq:ft_gt_expansion_E6xSU3}.
Since none of the subsequent tunings to obtain the type IV $\mathfrak{su}_3$ affect $f_2$, the factorization should also persist in the result \eqref{eq:E6xSU3_withU1_wrong}.

The factorization of $f_2 \rightarrow f_1^2$ reduces the number of free parameters in \eqref{eq:E6xSU3_withU1_wrong} by one.
Another one can be removed by noticing that the parameter $\lambda$ can be absorbed by a rescaling of $(u,v) \rightarrow \lambda^{-1/6} (u,v)$ and a redefinition of $f_1 \rightarrow f_1\,\lambda^{2/3}$.
With that, the resulting fibration
\begin{align}\label{eq:E6xSU3_withU1_final}
	\begin{split}
		f & = f_1^2 \, u^3 \, v^3 \, (u+p_3^2\,v)^2 \, , \\
		g &= \frac{1}{12} \,p_3^4\, u^4 \, v^6 \, (u+ p_3^2\,v)^2 \,   \, , \\
		\Delta & = \frac{1}{16} \, u^8 \, v^9 \, (u+ p_3^2\,v)^4 \, \left( 64 \, f_1^6 \, u \, (u+p_3^2\,v)^2 + 3 \,p_3^8 \, v^2   \right) \, ,
	\end{split}
\end{align}
now has the right number of singlets.
Therefore, F-theory compactified on \eqref{eq:E6xSU3_withU1_final} should be dual to the heterotic $\mathbb{Z}_3$-orbifold with the spectrum \eqref{eq:E6xSU3_spectrum_heterotic}.
Just like in the $\mathfrak{e}_7 \times \mathfrak{su}_2$ case, the single $({\bf 27,3})$ hypermultiplet is spread across three points on the base.
Hence, their appearance cannot be just a local phenomenon and should be further investigated in future work.

As a final consistency check, let us comment on the global structure of the gauge group of this model.
From the spectrum \eqref{eq:E6xSU3_spectrum_heterotic}, we see that the $\mathfrak{u}(1)$ charges of the matter states are tied to their non-abelian representations in a particular way:
\begin{itemize}
	\item The $\mathfrak{u}(1)$ is normalized such that the singlet has charge 1.
	\item Fundamentals under $\mathfrak{e}_7$ have half-integer charges.
	\item Fundamentals under $\mathfrak{e}_6$ and $\mathfrak{su}_3$ have charges in multiples of $\frac{1}{3}$; in the case of triplets, the charges are $\frac{2}{3} \! \! \mod \mathbb{Z}$.
	\item The bifundamental $({\bf 27,3)}$ has integer charge.
\end{itemize}
The associated gauge group thus has to be
\begin{align}
	\frac{E_6 \times SU(3) \times E_7 \times U(1)}{ \mathbb{Z}_6} = \frac{E_6 \times SU(3) \times E_7 \times U(1)}{ \mathbb{Z}_3 \times \mathbb{Z}_2} \, .
\end{align}
In F-theory, such non-trivial global gauge group structures are realized whenever the zero section and independent section intersect the codimension one fibers over the gauge divisor at different components in the resolved geometry \cite{Cvetic:2017epq}.\footnote{In this reference, the derivation of the global gauge group structure assumes a smooth global geometry, which we have not provided here.
Given the presence of the non-minimal point, a Calabi--Yau resolution of the tuned Weierstrass model without blow-ups in the base will most likely include non-flat fibers.
However, we expect the statement of \cite{Cvetic:2017epq} to still hold, since the global group structure only depends on the interplay between the rational sections and codimension one fibers, which in our model are still of Kodaira--Tate form.
}
For the three gauge algebras at hand, there is (up to automorphisms of the respective Dynkin diagrams)
only one possibility, which all shrink in the singular limit.
Thus, the independent section has to pass through the singular point over the gauge divisors.
Since in the model \eqref{eq:E6xSU3_withU1_final}, both $f$ and $g$ vanish over these codimension one loci ($\{u\}, \{v\}, \{u + p_3\,v^3\}$), the singular point of the elliptic fiber sits and $x = y =0$, which is indeed intersected by the section $(x,y) = (0, \sqrt{g})$.
Therefore, we see that our tuned F-theory geometry also accommodates the global gauge group structure of the heterotic orbifold.

\subsection{Adding More Bifundamentals}

Much as in the case of the $\mathfrak{e}_{7}\times\mathfrak{su}_2$ model, we
ask whether it is possible to engineer $\mathfrak{e}_{6}\times\mathfrak{su}%
_{3}$ gauge theories with more than one \textquotedblleft
exotic\textquotedblright\ bifundamental. Locally, at least, there seems to be
little issue with this procedure, provided we relax the constraints on the
self-intersection numbers of our curves. For example, we can simply add more
zeroes to produce additional delocalized bifundamentals.

Focussing on the purely gauge theoretic aspects of 6D anomaly cancellation,
the relevant constraints are (see e.g. \cite{Bershadsky:1996nh, Sadov:1996zm}):%
\begin{align}
\begin{split}
\left(  D_{\mathfrak{e}_{6}}\cdot D_{\mathfrak{e}_{6}}+6\right)   &
=N_{(\mathbf{27},\mathbf{1})}+3\times N_{(\mathbf{27},\mathbf{3})}\\
\left(  6D_{\mathfrak{su}(3)}\cdot D_{\mathfrak{su}(3)}+18\right)   &
=N_{(\mathbf{1},\mathbf{3})}+27\times N_{(\mathbf{27},\mathbf{3})}\\
D_{\mathfrak{e}_{6}}\cdot D_{\mathfrak{su}(3)}  &  =6\times N_{(\mathbf{27}%
,\mathbf{3})}.
\end{split}
\end{align}
where here, $D_{\mathfrak{g}}$ denotes the divisor class of a curve wrapped by
a seven-brane with gauge algebra $\mathfrak{g}$, and the $N$'s indicate the
multiplicity of various representations.

Increasing the number of bifundamentals necessarily raises the
self-intersection number relative to the case of a single bifundamental. As an
example with more than one bifundamental which satisfies these conditions, we
can take:%
\begin{align}
\begin{split}
	N_{(\mathbf{27},\mathbf{3})}  &  =2 \, , \quad N_{(\mathbf{27},\mathbf{1})}=0 \, , \quad N_{(\mathbf{1},\mathbf{3})}=0 \, , \\
	D_{\mathfrak{e}_{6}}\cdot D_{\mathfrak{e}_{6}}  &  =0 \, , \quad D_{\mathfrak{su}_{3}}\cdot D_{\mathfrak{su}(3)}=6 \, , \quad D_{\mathfrak{e}_{6}}\cdot D_{\mathfrak{su}(3)} =12 \, .
\end{split}
\end{align}
It seems plausible that this could arise in a compact geometry, but again, we
defer a full analysis to future work.

Much as in the case of $\mathfrak{e}_{7}\times\mathfrak{su}_2$ exotic
bifundamentals, we can compactify our 6D hypermultiplet on a genus $g$ curve,
and in so doing, produce $g$ 4D $\mathcal{N}=1$ chiral multiplets in the
$(\mathbf{27},\mathbf{3})$ representation of $\mathfrak{e}_{6}\times
\mathfrak{su}_3$, and $g$ in the conjugate representation. Again, this
discussion neglects subtleties having to do with embedding the matter curves
in a Calabi--Yau fourfold geometry, or the presence of ambient fluxes.

\section{Breaking Patterns \label{sec:BREAKING}}

\begin{figure}[t!]
\begin{center}
\scalebox{1}[1]{
\includegraphics[clip,scale=0.48]{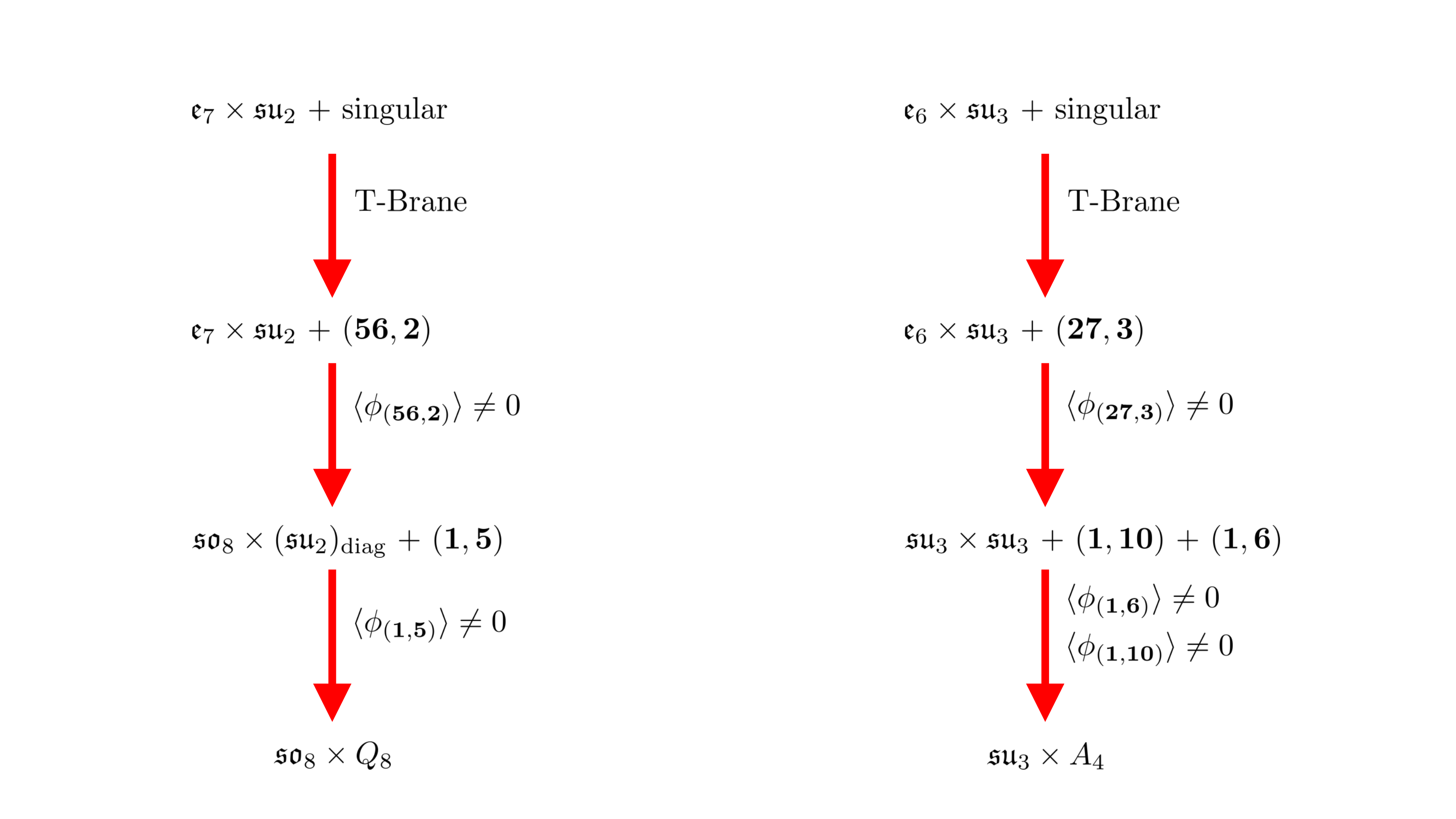}}
\end{center}
\caption{Depiction of breaking patterns associated with vevs for exotic bifundamentals.
Here, we assume that the continuous symmetries are flavor symmetries so that no
D-term constraints need to be satisfied. The final stage of the breaking patterns
leads to a finite order non-abelian group. Here, $Q_8$ denotes
the order eight quaternion group and $A_4$ the alternating group on four letters.
In the case of the breaking pattern for $\mathfrak{e}_6 \times \mathfrak{su}_3$ to $\mathfrak{su}_3 \times A_4$,
this also requires a vev for a scalar in the $(\mathbf{1},\mathbf{6})$ of $\mathfrak{su}_3 \times \mathfrak{su}_3$,
which descends from the $(\mathbf{27} , \mathbf{1})$ of $\mathfrak{e}_6 \times \mathfrak{su}_3$.}
\label{fig:BREAKING}
\end{figure}

In the previous sections, we focused on the geometric realization of exotic
bifundamental matter field. From this starting point, a number of additional
exotic structures can be realized in F-theory. The main idea is that by
activating background vevs for bifundamental matter fields, we can reach
unbroken symmetries with both high dimension representations of lower rank
symmetry groups, as well as finite non-abelian discrete symmetries. This follows
the general strategy advocated in reference \cite{Frampton:1994rk} to get non-abelian
finite groups from breaking patterns of a continuous symmetry group. See figure \ref{fig:BREAKING}
for a depiction of the breaking patterns associated with the exotic bifundamentals $(\mathbf{56},\mathbf{2})$ of $\mathfrak{e}_7 \times \mathfrak{su}_2$ and the $(\mathbf{27},\mathbf{3})$ of $\mathfrak{e}_6 \times \mathfrak{su}_3$.

Now, in the specific geometries we have constructed, we have a single
bifundamental hypermutiplet between two gauge groups factors. Activating a vev
in this case is obstructed because of the D-term constraints from the two
gauge algebras. The reason comes down to the fact that to actually Higgs a
$\mathfrak{su}_N \times\mathfrak{g}$ gauge theory with matter fields in the
fundamental representation, we need at least two hypermultiplets (see e.g
\cite{Morrison:2012np}).\footnote{To see why this is so, it is enough to consider
the super-Higgs mechanism for $\mathfrak{su}_N$ gauge theory. Suppose we have
a theory of $M$ hypermultiplets in the fundamental representation. Writing the
collection of scalars as pairs of complex scalars $q_{I}$ and $\widetilde{q}%
_{I}$ for $I=1,...,M$ such that each doublet $(q_{I},\widetilde{q}_{I}^{\dag
})$ rotates as a doublet under $SU(2)$ R-symmetry, a breaking pattern of
$\mathfrak{su}_N$ requires us to give a vev to two hypermultipets. Moreover,
as can be seen from the way the Goldstone bosons are eaten, we need one set of
complex scalars transforming in the fundamental representation, and one set of
complex scalars in the conjugate representation to get a vev. This is
problematic when we have a single bifundamental, since then, only $q$ or
$\widetilde{q}$ can be eaten.} Of course, we can take a decompactification
limit where all gauge group factors turn into flavor symmetry factors. In this
limit (which is also decoupled from gravity), there are no D-term constraints
to satisfy, and we can reduce our problem to a purely group theoretic
exercise. If we insist on a model where all symmetries are gauged, we can also
contemplate models with more than one exotic bifundamental. As we have
indicated in previous sections, there appears to be no a priori obstruction to
realizing 6D\ vacua with this feature, and even fewer constraints for 4D vacua.

With this in mind, our aim here will be to explore the breaking patterns
generated by giving a vev to an exotic bifundamental. For ease of exposition,
we focus on the special case where the pair of gauge algebras are decoupled,
namely we treat them as global symmetries. Note that since the exotic bifundamentals
are delocalized across multiple points of the geometry, we must also keep all such singular points localized
in a single patch of the geometry.

Of course, we can also weakly gauge these flavor symmetries
(and introduce a corresponding tensor multiplet to cancel
gauge anomalies) and obtain similar breaking patterns in models where there is
more than one exotic bifundamental. As we have already remarked, the explicit geometric
realization of these cases is likely to be more involved, though a priori, there does not
appear to be any bottom up obstruction to realizing such a model.

Our plan in the rest of this section will be to study the resulting breaking patterns from
the $(\mathbf{56},\mathbf{2})$ of $\mathfrak{e}_{7}\times\mathfrak{su}_2$ and
the $(\mathbf{27},\mathbf{3})$ of $\mathfrak{e}_{6}\times\mathfrak{su}_3$.

\subsection{The $(\mathbf{56} , \mathbf{2})$ of $\mathfrak{e}_7 \times \mathfrak{su}_2$}

As a first example, consider the breaking pattern generated by a single
$(\mathbf{56},\mathbf{2})$ of $\mathfrak{e}_{7}\times\mathfrak{su}_2$.
Returning to our discussion in Appendix \ref{app:E7SU2}, we observe that there is a
singlet in the decomposition to the subalgera $\mathfrak{so}_{8} \times (\mathfrak{su}_2)_{\text{diag}}$:
\begin{align*}
	\mathfrak{e}_{7}\times\mathfrak{su}_2  & \supset\mathfrak{so}_{8} \times (\mathfrak{su}_2)_{\text{diag}}\\
	(\mathbf{56},\mathbf{2})  & \rightarrow(\mathbf{8}_{v},\mathbf{3})+(\mathbf{8}_{s},\mathbf{3})+(\mathbf{8}_{c},\mathbf{3})\\
	& +(\mathbf{8}_{v},\mathbf{1})+(\mathbf{8}_{s},\mathbf{1})+(\mathbf{8}_{c},\mathbf{1})\\
	& +(\mathbf{1},\mathbf{5})+(\mathbf{1},\mathbf{3})^{\oplus3}+(\mathbf{1},\mathbf{1})^{\oplus2}.
\end{align*}
Note also that the coset space $\mathfrak{e}_{7}\times\mathfrak{su}_2 /\mathfrak{so}_{8}\times (\mathfrak{su}_2)_{\text{diag}}$ has dimension
$105$, the same as the Goldstone modes:%
\begin{align*}
\text{Goldstone Modes}  \quad & \text{: }(\mathbf{8}_{v},\mathbf{3})+(\mathbf{8}%
_{s},\mathbf{3})+(\mathbf{8}_{c},\mathbf{3})\\
& +(\mathbf{8}_{v},\mathbf{1})+(\mathbf{8}_{s},\mathbf{1})+(\mathbf{8}%
_{c},\mathbf{1})+(\mathbf{1},\mathbf{3})^{\oplus3}.
\end{align*}
Nevertheless, as we have already remarked, the D-flatness conditions cannot be
satisfied with a single exotic bifundamental. Taking this into acount, we see
that the leftover modes not parameterizing the coset includes the
$(\mathbf{1},\mathbf{5})$ of $\mathfrak{so}_{8}\times\mathfrak{su}%
(2)_{\text{diag}}$. This is a high-dimensional representation, and as we have
remarked, it has proven difficult to engineer string compactifications with
this representation, though as far as we are aware there is no
\textquotedblleft no-go\textquotedblright\ theorem forbidding the appearance
of these representations (see, however, \cite{Klevers:2017aku}). Indeed, the present
discussion suggests that it ought to be possible to find models with more than
one exotic bifundamental, and thus higgs further to such an exotic representation.

This also provides a way for us to generate a high order non-abelian discrete
group. To see why, we note that a vev for the $\mathbf{5}$ of $(\mathfrak{su}_2)_{\text{diag}}$ leaves unbroken the order eight group known as $Q_{8}$, the
quaternion group \cite{Adulpravitchai:2009kd}.\footnote{Recall that $Q_{8}$ is the
group with relations $\left\langle \overline{e},i,j,k|\overline{e}^{2}%
=e,i^{2}=j^{2}=k^{2}=ijk=\overline{e}\right\rangle $, with $e$ the identity.
This group has order eight.}

\subsection{The $(\mathbf{27} , \mathbf{3})$ of $\mathfrak{e}_6 \times \mathfrak{su}_3$}

As another example, consider the breaking pattern generated by a single
$(\mathbf{27},\mathbf{3})$ of $\mathfrak{e}_{6}\times\mathfrak{su}_3$.
Returning to our discussion in Appendix \ref{app:E6SU3}, we observe that there is a
singlet in the decomposition to the subalgebra $(\mathfrak{su}_{3})_{(3)}%
\times(\mathfrak{su}_{3})_{\text{diag}}$:%
\begin{align*}
\mathfrak{e}_{6}\times\mathfrak{su}_3  & \supset(\mathfrak{su}_{3}%
)_{(3)}\times(\mathfrak{su}_{3})_{\text{diag}}\\
(\mathbf{27},\mathbf{3})  & \rightarrow(\mathbf{1},\mathbf{8})^{\oplus
2}+(\mathbf{1},\mathbf{1})+(\mathbf{1},\mathbf{10})+(\mathbf{3},\mathbf{8}%
)+(\mathbf{3},\mathbf{1})+(\overline{\mathbf{3}},\mathbf{8})+(\overline
{\mathbf{3}},\mathbf{1}).
\end{align*}
Note also that the coset space $\mathfrak{e}_{6}\times\mathfrak{su}_3/(\mathfrak{su}_{3})_{(3)}\times(\mathfrak{su}_{3})_{\text{diag}}$ has
dimension $70$, the same as the Goldstone modes:%
\[
\text{Goldstone Modes}\text{: }(\mathbf{1},\mathbf{8})^{\oplus2}%
+(\mathbf{3},\mathbf{8})+(\mathbf{3},\mathbf{1})+(\overline{\mathbf{3}%
},\mathbf{8})+(\overline{\mathbf{3}},\mathbf{1}),
\]
which leaves us with a matter field in the $\mathbf{10}$ of $\mathfrak{su}_3$.
This is the three-index symmetric representation of $\mathfrak{su}_3$,
and has been viewed as difficult to engineer in F-theory \cite{Klevers:2017aku}.
Here, we see that if there really is an issue, it cannot be ascribed to local
geometric conditions but must instead have to do with subtle global questions
about compact Calabi--Yau threefolds.

Giving a further vev to this uneaten mode results in a continuous group
\cite{Luhn:2011ip}, but in tandem with a vev for matter in the $\mathbf{6}$ of
$\mathfrak{su}_3$, all continuous symmetries are broken. Note that the
decomposition of the fundamental of $\mathfrak{e}_{6}$ contains such a
representation:%
\begin{align*}
\mathfrak{e}_{6}\times\mathfrak{su}_3  & \supset(\mathfrak{su}_{3}%
)_{(3)}\times(\mathfrak{su}_{3})_{\text{diag}}\\
(\mathbf{27},\mathbf{1})  & \rightarrow(\mathbf{1},\mathbf{6})+(\mathbf{1}%
,\overline{\mathbf{3}})+(\mathbf{3},\overline{\mathbf{3}})+(\overline
{\mathbf{3}},\overline{\mathbf{3}}).
\end{align*}
The combination of a vev in the $\mathbf{10}$ and $\mathbf{6}$ of
$\mathfrak{su}_3$ breaks the continuous group to $A_{4}$ the
alternating group on four letters \cite{Luhn:2011ip}. Note that we typically have many
matter fields in the $(\mathbf{27},\mathbf{1})$. From this perspective, the
primary challenge in realizing compact models will, as stated before, be to
obtain models with more than one exotic bifundamental.

\section{Conclusions \label{sec:CONC}}

F-theory is a remarkably flexible framework for realizing a broad class of low energy effective field theories
via string compactification. From this perspective, it is important to understand the list of possible exotic phenomena
involving singular limits of moduli in such backgrounds. Such singular limits can lead to novel structures in low energy effective field
theory, including high dimension representations for matter multiplets, as well as non-abelian discrete symmetries.
In this paper we have taken some preliminary steps in developing a systematic framework for the construction of ``exotic'' bifundamental matter,
as well as the subsequent symmetry breaking patterns generated by such matter fields. From a geometric perspective, the main idea in our approach
is to allow the appearance of singular elliptically fibered Calabi--Yau threefolds in which the multiplicity of vanishing at a
collision of two components of the discriminant locus is at least a $(4,6,12)$ point. When this collision has a sufficiently high
order of tangency, there are additional directions in the Higgs branch of moduli space, and it is possible to activate a T-brane deformation,
moving away from the SCFT point, and instead to a weakly coupled gauge theory. This new point in moduli space is described by the same
singular Calabi--Yau geometry, but is non-singular in the physics description. We have illustrated these general considerations in the
special case of bifundamental matter for $\mathfrak{e}_7 \times \mathfrak{su}_2$ and $\mathfrak{e}_6 \times \mathfrak{su}_3$, matching them to
explicit dual heterotic orbifold models. Additionally, further higgsing of these theories leads to matter in ``exotic'' high dimension representations of $\mathfrak{su}_2$ and $\mathfrak{su}_3$. In the remainder of this section we discuss some avenues of further investigation.

One of the items left open by our analysis is a general criterion for the order of tangency needed in order to activate a T-brane deformation to a weakly coupled gauge theory. Rather, we have used a combination of top down and bottom up considerations to provide plausible candidates for such a match. In particular, we have often relied on some global features of heterotic/F-theory duality, even though the appearance of exotic bifundamental matter appears to be an issue which is localized in the geometry of an F-theory compactification.
Developing a concise description of general necessary and sufficient conditions to realize such exotic bifundamental matter
would seem a worthwhile undertaking.
Specifically, the feature that the bifundamental hypermultiplet is spread across several points deserves further investigation.
A better understanding of this phenomenon could provide us with methods to construct duals of other orbifolds with interesting matter (e.g., $\mathbb{Z}_3$-orbifolds with the three-index antisymmetric (${\bf 84}$) of $\mathfrak{su}_9$), and to extend our analysis to F-theory vacua which do not have heterotic duals.

We have also presented evidence that we can engineer discrete non-abelian symmetries in F-theory by performing an appropriate higgsing with such
exotic bifundamental matter. We have deferred an explicit construction of models with more than one hypermultiplet in a bifundamental
representation, but as we have already remarked, there does not appear to be any issue with engineering such examples from the perspective
of bottom up considerations. Further, a geometric characterization of non-abelian discrete symmetries in F-theory remains an important open problem to understand better.

Along these lines, it would be interesting to better understand in systematic terms the class of discrete symmetries which can actually be realized in string compactification. The present method using exotic bifundamental matter provides some examples, but it would be interesting to further analyze whether there are sharp upper bounds on the order, structure and representations of such delicate objects.

Another important direction to explore further is the behavior of such exotic bifundamental matter upon compactification to four dimensions.
One of the standard tools for understanding chiral matter spectra in particular involves coupling hypermultiplets localized on matter curves to
various gauge field and bulk fluxes. Due to the way these matter fields are engineered in F-theory, it would be interesting to examine
the conditions necessary to realize a chiral matter spectrum for them. Turning the discussion around, this may also provide insight into non-perturbative corrections to heterotic orbifold compactifications.

\section*{Acknowledgements}

We thank F.~Apruzzi, C.~Lawrie, and T.~Rudelius for helpful discussions.
JJH and LL thank the Banff International Research
Station for hospitality during workshop 18w5190 on the Geometry and Physics of F-theory.
The work of MC and LL is supported by DOE Award DE-SC0013528. MC further acknowledges
the support by the Fay R.~and Eugene L.~Langberg Endowed Chair and the Slovenian Research Agency.
The work of JJH is supported by NSF CAREER grant PHY-1756996.

\appendix

\section{Tate Model with $\mathfrak{e}_6 \times \mathfrak{su}_3$ \label{app:Tate_E6xSU3}}

In this Appendix, we present the generic Tate model of an elliptic fibration over $B = \bbF_6$, whose codimension one singularities correspond to an $\mathfrak{e}_6 \times \mathfrak{su}_3 \times \mathfrak{e}_6$ gauge algebra in F-theory. To begin with, any Weierstrass model over $\mathbb{F}_6$ has a non-higgsable $\mathfrak{e}_6$ on the $-6$-curve $\{v\}$ due to the factorization
\begin{align}
	f = v^3 \, \ft \, ,  \quad g = v^4 \, \gt \, ,
\end{align}
where $[\ft] = 5[u] + 2[s] = 5[v] + 32[s]$ and $[\gt] = 8[u] = 8[v] + 48[s]$.
Let $\sigma=0$ denote the locus of the $\mathfrak{su}_3$, with $[\sigma] = [u]$ a $+6$-curve.
Expanding
\begin{align}\label{eq:ansatz_1_E6SU3}
	\begin{split}
		\ft & = f_{32}\,v^5 + f_{26}\,v^4\,\sigma + f_{20}\,v^3\,\sigma^2 + \sigma^3\,f' \, , \\
		\gt & = g_{48}\,v^8 + g_{42}\,v^7\,\sigma + g_{36}\,v^6\,\sigma^2 + \sigma^3\,g' \, ,
	\end{split}
\end{align}
the generic way to obtain a split I$_3$ fiber over $\sigma=0$ is to set
\begin{align}
	f_{32} = -3\,\phi_8^4 \, , \quad g_{48}=2\,\phi_8^6 \, , \quad g_{42} = -\phi_{18}\,\phi_8^3 \, , \quad f_{26} = \phi_{18}\,\phi_8 \, , \quad g_{36} = \frac{\phi_{18}^2 - 12\,f_{20} \,\phi_8^2}{12} \, .
\end{align}
Next, we want to tune a split type IV fiber on another +6-curve.
Follow the $\mathfrak{e}_7 \times \mathfrak{su}_2$ example, we label this curve by $u=0$, and the $\mathfrak{su}_3$ curve by $\sigma \equiv u + p_6\,v =0$ for a generic degree 6 polynomial $p_6$.
For that, we have to tune the remaining coefficients in \eqref{eq:ansatz_1_E6SU3} such that $\ft \sim u^3$ and $\gt \sim u^4$.
By degree counting, the general expansion for the primed functions are
\begin{align}
	\begin{split}
		f' & = f_2 \, u^2 + f_8\,u\,v + f_{14}\,v^2 \, , \\
		g' &= g_0\,u^5 + g_6\,u^4\,v + g_{12}\,u^3\,v^2 + g_{18} \, u^2 \, v^3 + g_{24}\,u\,v^4 +  g_{30} \, v^5
	\end{split}
\end{align}
Inserting this into \eqref{eq:ansatz_1_E6SU3}, together with $\sigma = u+p_6\,v$, we can again solve the vanishing order conditions using standard UFD methods.

The tricky part is implementing the split condition for $\mathfrak{e}_6$.
Explicitly, this requires the following function to be a square:
\begin{align}
	(\tilde{g}/u^4)|_{u=0} = 12\,g_6\,p_6 + 12\,f_2\,c_2^2\,p_6 + c_6^2 \, .
\end{align}
The most general way to do it is by setting
\begin{align}
	g_6 = \frac{1}{12} \left( \alpha^2 \, p_6 -12\,f_2\,c_2^2 + 2\,\alpha \, c_6 \right) \, ,
\end{align}
where $\alpha$ is an arbitrary complex number.
The final result of the tuning process is
\begin{align}\label{eq:weierstrass_functions_E6_Tate}
	\begin{split}
		f  = \, & \left [f_2\, u^2 + (2\, f_2\, p_6 - 3\, c_2^4 + c_2\, c_6 )\, u\, v + p_6\, (f_2\, p_6 - c_2\, c_6)\, v^2 \right] \, u^3 \, v^3\, , \\
		g =\, & \left[ 12\, g_0\,u^4  +  \left( 36 \,g_0\,p_6 + 2\,\alpha\,c_6 + \alpha^2\,p_6 - 12\,f_2\,c_2^2 \right) u^3\,v   \right. \\
       			&  + \left( 3\,(12\,g_0+\alpha^2) \,p_6^2 + 24\,c_2^2\,(c_2^4 - f_2\,p_6) + c_6 \, (c_6 + 12\,c_2^3 + 6\,\alpha\,p_6 ) \right)  u^2\,v^2 \\
			  & + \left( (12\, g_0 + 3\,\alpha^2)\,p_6^2 + 2\,c_6 \, (6\,c_2^3 + c_6) + 6\,p_6 \, (\alpha\,c_6 - 2\,f_2\,c_2^2)  \right)  \,p_6\,u\,v^3   \\
      		&    \left. + p_6^2\,( c_6  + p_6\,\alpha)^2 \,v^4 \right] \, \frac{u^4 \, v^4}{12} \, , \\
		\Delta = \, & u^8\,v^8\,(u+p_6\,v)^3 \, \Delta_{\text{res}} \, ,
	\end{split}
\end{align}
with
\begin{align}\label{eq:res_disc_E6_Tate}
\begin{split}
	\Delta_{\text{res}} & = 27\,g_0^2\,u^5 + (...) \, u^4v + (...)\,u^3v^2 + (...) \, u^2v^3 + (...)\, u\,v^4 + \frac{p_6\,(\alpha \, p_6+ c_6)^4}{144}\, v^5 \\
	&= 27\,g_0^2\,s^5 + (...)\,s^4 v + (...)\,s^3v^2 + (...)\,s^2 v^3 + (...)\,s\,v^4 + p_6\,c_2^3\,Q_{18}\,v^5 \, .
\end{split}
\end{align}

After blowing up the non-minimal points at $u=p_6=0$ and thus introducing six tensors, both the $\mathfrak{e}_6$ and $\mathfrak{su}_3$ divisors have self-intersection 0.
Consistent with anomalies, the intersections with the residual discriminant result in six points of $\mathfrak{e}_7$ enhancement for $u=0$ with ${\bf 27}$ matter of $\mathfrak{e}_6$, and 18 points of I$_4$ enhancement for $\sigma = 0$ with ${\bf 3}$ matter of $\mathfrak{su}_3$.
The free parameters of the Weierstrass model \eqref{eq:weierstrass_functions_E6_Tate} are contained in $f_2, c_2, c_6, p_6,g_0, \alpha$, which after subtracting $3+1$ reparametrization degree of freedoms yield 18 uncharged hypers.
One can verify this counting by checking the gravitational anomaly, which is indeed canceled:
\begin{align}
	\begin{split}
		H & = 6 \times 27 + 18 \times 3 + 18 = 234 \, , \\
		V & = 2 \times 78 + 8 = 164 \, , \\
		T & = 7 \, \\
		\Longrightarrow \quad 273 & = H - V + 29T \, .
	\end{split}
\end{align}

\section{Branching Rules for Bifundamental Multiplets}

In this Appendix we review some aspects of branching rules for representations
of $\mathfrak{e}_{7}\times\mathfrak{su}_2$ and $\mathfrak{e}_6 \times \mathfrak{su}_3$. This is used in the main body of
the text to study the appearance of higher dimension representations of
$\mathfrak{su}_2$ and $\mathfrak{su}_3$. All of this material is standard, and can be extracted for
example from \cite{mckay1981tables} as well as computer packages such as
\texttt{LieART} \cite{Feger:2012bs}.

\subsection{$\mathfrak{e}_7 \times \mathfrak{su}_2$ \label{app:E7SU2}}

To begin, we observe the following branching rules for representations of
$\mathfrak{e}_{7}$ to some subalgebras. We begin with the decomposition to
$\mathfrak{so}_{12}\times\mathfrak{su}_{2}$.%
\begin{align}
\mathfrak{e}_{7}  &  \supset\mathfrak{so}_{12}\times\mathfrak{su}_{2}\\
\mathbf{56}  &  \rightarrow(\mathbf{12},\mathbf{2})+(\mathbf{32},\mathbf{1})\\
\mathbf{133}  &  \rightarrow(\mathbf{66},\mathbf{1})+(\mathbf{32}%
,\mathbf{2})+(\mathbf{1},\mathbf{3}).
\end{align}
It is also helpful to consider the decomposition of $\mathfrak{so}_{12}$ to
the subalgebra $\mathfrak{so}_{8}\times\mathfrak{su}_{2}\times\mathfrak{su}%
_{2}$:
\begin{align}
\mathfrak{so}_{12}  &  \supset\mathfrak{so}_{8}\times\mathfrak{su}_{2}%
\times\mathfrak{su}_{2}\\
\mathbf{12}  &  \rightarrow(\mathbf{8}_{v},\mathbf{1},\mathbf{1}%
)+(\mathbf{1},\mathbf{2},\mathbf{2})\\
\mathbf{32}  &  \rightarrow(\mathbf{8}_{s},\mathbf{2},\mathbf{1}%
)+(\mathbf{8}_{c},\mathbf{1},\mathbf{2})\\
\mathbf{66}  &  \rightarrow(\mathbf{28},\mathbf{1},\mathbf{1})+(\mathbf{8}%
_{v},\mathbf{2},\mathbf{2})+(\mathbf{1},\mathbf{3},\mathbf{1})+(\mathbf{1}%
,\mathbf{1},\mathbf{3}),
\end{align}
where in the above, $\mathbf{8}_{s}$, $\mathbf{8}_{c}$ denote the two spinor
representations of $\mathfrak{so}_{8}$.

Putting this together, we now consider the decompositions of $\mathfrak{e}%
_{7}\times\mathfrak{su}_{2}$ to the subalgebra $(\mathfrak{so}_{8}%
\times\left(  \mathfrak{su}_{2}\right)  ^{3})\times\mathfrak{su}_{2}$, in the
obvious notation:%
\begin{align}
\mathfrak{e}_{7}\times\mathfrak{su}_{2}  &  \supset\mathfrak{so}_{8}%
\times\left(  \mathfrak{su}_{2}\right)  ^{3}\times\mathfrak{su}_{2}\\
(\mathbf{56},\mathbf{2})  &  \rightarrow(\mathbf{8}_{v},\mathbf{1}%
,\mathbf{1},\mathbf{2},\mathbf{2})+(\mathbf{1},\mathbf{2},\mathbf{2}%
,\mathbf{2},\mathbf{2})+(\mathbf{8}_{s},\mathbf{2},\mathbf{1},\mathbf{1}%
,\mathbf{2})+(\mathbf{8}_{c},\mathbf{1},\mathbf{2},\mathbf{1},\mathbf{2})\\
(\mathbf{56},\mathbf{1})  &  \rightarrow(\mathbf{8}_{v},\mathbf{1}%
,\mathbf{1},\mathbf{2},\mathbf{1})+(1,\mathbf{2},\mathbf{2},\mathbf{2}%
,\mathbf{1})+(\mathbf{8}_{s},\mathbf{2},\mathbf{1},\mathbf{1},\mathbf{1}%
)+(\mathbf{8}_{c},\mathbf{1},\mathbf{2},\mathbf{1},\mathbf{1})\\
(\mathbf{133},\mathbf{1})  &  \rightarrow(\mathbf{28},\mathbf{1}%
,\mathbf{1},\mathbf{1},\mathbf{1})\\
&  +(\mathbf{8}_{v},\mathbf{2},\mathbf{2},\mathbf{1},\mathbf{1})+(\mathbf{8}%
_{s},\mathbf{2},\mathbf{1},\mathbf{2},\mathbf{1})+(\mathbf{8}_{c}%
,\mathbf{1},\mathbf{2},\mathbf{2},\mathbf{1})\\
&  +(\mathbf{1},\mathbf{3},\mathbf{1},\mathbf{1},\mathbf{1})+(\mathbf{1}%
,\mathbf{1},\mathbf{3},\mathbf{1},\mathbf{1})+(\mathbf{1},\mathbf{1}%
,\mathbf{1},\mathbf{3},\mathbf{1})\\
(\mathbf{1},\mathbf{2})  &  \rightarrow(\mathbf{1},\mathbf{1},\mathbf{1}%
,\mathbf{1},\mathbf{2})\\
(\mathbf{1},\mathbf{3})  &  \rightarrow(\mathbf{1},\mathbf{1},\mathbf{1}%
,\mathbf{1},\mathbf{3}).
\end{align}

As per our discussion of breaking patterns in section \ref{sec:BREAKING}, we seek out the
branching rules to the diagonal subalgebra $\mathfrak{so}_{8}\times\left(
\mathfrak{su}_{2}\right)  _{\text{diag}}$:%
\begin{align}
\mathfrak{e}_{7}\times\mathfrak{su}_{2}  &  \supset\mathfrak{so}_{8}%
\times\left(  \mathfrak{su}_{2}\right)  _{\text{diag}}\\
(\mathbf{56},\mathbf{2})  &  \rightarrow(\mathbf{8}_{v},\mathbf{3}%
)+(\mathbf{8}_{s},\mathbf{3})+(\mathbf{8}_{c},\mathbf{3})\\
&  +(\mathbf{8}_{v},\mathbf{1})+(\mathbf{8}_{s},\mathbf{1})+(\mathbf{8}%
_{c},\mathbf{1})\\
&  +(\mathbf{1},\mathbf{5})+(\mathbf{1},\mathbf{3})^{\oplus3}+(\mathbf{1}%
,\mathbf{1})^{\oplus2}\\
(\mathbf{56},\mathbf{1})  &  \rightarrow(\mathbf{8}_{v},\mathbf{2}%
)+(\mathbf{8}_{s},\mathbf{2})+(\mathbf{8}_{c},\mathbf{2})+(\mathbf{1}%
,\mathbf{4})+(\mathbf{1},\mathbf{2})^{\oplus2}\\
(\mathbf{133},\mathbf{1})  &  \rightarrow(\mathbf{28},\mathbf{1})\\
&  +(\mathbf{8}_{v},\mathbf{3})+(\mathbf{8}_{s},\mathbf{3})+(\mathbf{8}%
_{c},\mathbf{3})\\
&  +(\mathbf{8}_{v},\mathbf{1})+(\mathbf{8}_{s},\mathbf{1})+(\mathbf{8}%
_{c},\mathbf{1})\\
&  +(\mathbf{1},\mathbf{3})+(\mathbf{1},\mathbf{3})+(\mathbf{1},\mathbf{3})\\
(\mathbf{1},\mathbf{2})  &  \rightarrow(\mathbf{1},\mathbf{2})\\
(\mathbf{1},\mathbf{3})  &  \rightarrow(\mathbf{1},\mathbf{3}).
\end{align}

\subsection{$\mathfrak{e}_6 \times \mathfrak{su}_3$ \label{app:E6SU3}}


Consider now the decompositions of representations for
$\mathfrak{e}_{6}\times\mathfrak{su}_{3}$. To begin, we observe the following
branching rules for representations of $\mathfrak{e}_{6}$ to some subalgebras:%
\begin{align}
\mathfrak{e}_{6}  &  \supset\mathfrak{su}_{3}\times\mathfrak{su}_{3}%
\times\mathfrak{su}_{3}\\
\mathbf{27}  &  \rightarrow(\mathbf{3},\mathbf{3},\mathbf{1})+(\overline
{\mathbf{3}},\mathbf{1},\mathbf{3})+(\mathbf{1},\overline{\mathbf{3}%
},\overline{\mathbf{3}})\\
\mathbf{78}  &  \rightarrow(\mathbf{3},\overline{\mathbf{3}},\mathbf{3}%
)+(\overline{\mathbf{3}},\mathbf{3},\overline{\mathbf{3}})+(\mathbf{8}%
,\mathbf{1},\mathbf{1})+(\mathbf{1},\mathbf{8},\mathbf{1})+(\mathbf{1}%
,\mathbf{1},\mathbf{8}).
\end{align}
The decomposition of the various representations of $\mathfrak{e}_{6}%
\times\mathfrak{su}_{3}$ to $(\mathfrak{su}_{3})^{3}\times\mathfrak{su}_{3}$
is then:%
\begin{align}
\mathfrak{e}_{6}\times\mathfrak{su}_{3}  &  \supset\left(  \mathfrak{su}%
_{3}\right)  ^{3}\times\mathfrak{su}_{3}\\
(\mathbf{27},\mathbf{3})  &  \rightarrow(\mathbf{3},\mathbf{3},\mathbf{1}%
,\mathbf{3})+(\overline{\mathbf{3}},\mathbf{1},\mathbf{3},\mathbf{3}%
)+(\mathbf{1},\overline{\mathbf{3}},\overline{\mathbf{3}},\mathbf{3})\\
(\mathbf{27},\mathbf{1})  &  \rightarrow(\mathbf{3},\mathbf{3},\mathbf{1}%
,\mathbf{1})+(\overline{\mathbf{3}},\mathbf{1},\mathbf{3},\mathbf{1}%
)+(\mathbf{1},\overline{\mathbf{3}},\overline{\mathbf{3}},\mathbf{1})\\
(\mathbf{78},\mathbf{1})  &  \rightarrow(\mathbf{3},\overline{\mathbf{3}%
},\mathbf{3},\mathbf{1})+(\overline{\mathbf{3}},\mathbf{3},\overline
{\mathbf{3}},\mathbf{1})+(\mathbf{8},\mathbf{1},\mathbf{1},\mathbf{1}%
)+(\mathbf{1},\mathbf{8},\mathbf{1},\mathbf{1})+(\mathbf{1},\mathbf{1}%
,\mathbf{8},\mathbf{1})\\
(\mathbf{1},\mathbf{3})  &  \rightarrow(\mathbf{1},\mathbf{1},\mathbf{1}%
,\mathbf{3})\\
(\mathbf{1},\mathbf{8})  &  \rightarrow(\mathbf{1},\mathbf{1},\mathbf{1}%
,\mathbf{8}).
\end{align}
We now single out the third $\left(  \mathfrak{su}_{3}\right)  _{(3)}$ factor,
and take the diagonal subalgebra of the first, second and fourth factors. The
various representations then decompose as:%
\begin{align}
\mathfrak{e}_{6}\times\mathfrak{su}_{3}  &  \supset\left(  \mathfrak{su}%
_{3}\right)  _{(3)}\times\left(  \mathfrak{su}_{3}\right)  _{\text{diag}}\\
(\mathbf{27},\mathbf{3})  &  \rightarrow(\mathbf{1},\mathbf{8})^{\oplus
2}+(\mathbf{1},\mathbf{1})+(\mathbf{1},\mathbf{10})+(\mathbf{3},\mathbf{8}%
)+(\mathbf{3},\mathbf{1})+(\overline{\mathbf{3}},\mathbf{8})+(\overline
{\mathbf{3}},\mathbf{1})\\
(\mathbf{27},\mathbf{1})  &  \rightarrow(\mathbf{1},6)+(\mathbf{1}%
,\overline{\mathbf{3}})+(\mathbf{3},\overline{\mathbf{3}})+(\overline
{\mathbf{3}},\overline{\mathbf{3}})\\
(\mathbf{78},\mathbf{1})  &  \rightarrow(3,\mathbf{8})+(\mathbf{3}%
,\mathbf{1})+(\overline{\mathbf{3}},\mathbf{8})+(\overline{\mathbf{3}%
},\mathbf{1})+(\mathbf{1},\mathbf{8})^{\oplus2}+(\mathbf{8},\mathbf{1})\\
(\mathbf{1},\mathbf{3})  &  \rightarrow(\mathbf{1},\mathbf{3})\\
(\mathbf{1},\mathbf{8})  &  \rightarrow(\mathbf{1},\mathbf{8}).
\end{align}

\newpage

\bibliographystyle{utphys}
\bibliography{bifundos}

\end{document}